# Beyond Greenfield: The D³ Framework for AI-Driven Productivity in Brownfield Engineering

**Krishna Kumaar Sharma**[1]
Formerly **Head of Engineering - AI, Amazon**
krishna.sharma1408@gmail.com | linkedin.com/in/krisberlin

## Abstract

Brownfield engineering work involving legacy systems, incomplete documentation, and fragmented architectural knowledge poses unique challenges for the effective use of large language models (LLMs). Prior research has largely focused on greenfield or synthetic tasks, leaving a gap in structured workflows for complex, context-heavy environments. This paper introduces the Discover–Define–Deliver (D³) Framework, a disciplined LLM-assisted workflow that combines role-separated prompting strategies with applied best practices for navigating ambiguity in brownfield systems. The framework incorporates a dual-agent prompting architecture in which a Builder model generates candidate outputs and a Reviewer model provides structured critique to improve reliability. I conducted an exploratory survey study with 52 software practitioners who applied the D³ workflow to real-world engineering tasks such as legacy system exploration, documentation reconstruction, and architectural refactoring. Respondents reported perceived improvements in task clarity, documentation quality, and cognitive load, along with self-estimated productivity gains. In this exploratory study, participants reported a weighted average productivity improvement of 26.9%, reduced cognitive load for approximately 77% of participants, and 83% of participants spent less time fixing or rewriting code due to better initial planning with AI. As these findings are self-reported and not derived from controlled experiments, they should be interpreted as preliminary evidence of practitioner sentiment rather than causal effects. The results highlight both the potential and limitations of structured LLM workflows for legacy engineering systems and motivate future controlled evaluations.

---

# 1. Introduction: How AI Drives Productivity in Brownfield Environments

## 1.1 The Productivity Bottlenecks in Large, Legacy Systems

Large enterprises face a fundamental productivity paradox: while AI coding tools promise dramatic efficiency gains, their impact in brownfield environments characterized by legacy codebases, complex dependencies, and accumulated technical debt remains inconsistent and often disappointing. Recent industry surveys indicate widespread AI adoption in software development. The 2025 DORA report found that 90% of survey respondents use AI at work (Harvey et al.), while McKinsey research demonstrated that developers can complete tasks up to twice as fast with generative AI tools (Deniz et al.). A brownfield codebase typically spanning millions of lines, built across multiple platforms, maintained by dozens of teams, and evolved over years or decades embodies enormous business value and accumulated domain knowledge. Yet this same complexity imposes severe cognitive and coordination costs on anyone working within it.

Research into developer productivity has identified several persistent bottlenecks in large organizations. First, **fragmented knowledge** manifests as scattered understanding: design rationales live in archived tickets, architectural decisions exist only in the heads of departed engineers, and critical business constraints are embedded in comments or institutional memory rather than documented systematically. When an engineer needs to understand why a particular module exists or how it interacts with seven downstream services, they face hours of exploration, trial-and-error debugging, and informal inquiry. Second, **weak or outdated documentation** creates recurring friction. Many teams lack coherent runbooks, API documentation, or design decision records; when documentation does exist, it often diverges from reality, forcing engineers to rely on code reading and experimentation to learn the ground truth. Third, **dependency complexity** makes even simple changes risky and costly. Changing a core utility or data structure can affect hundreds of call sites across the codebase; engineers must mentally trace impact graphs, reason about invariants, and coordinate with multiple teams to avoid breaking changes. Fourth, **coordination overhead** multiplies as teams scale. Code reviews become bottlenecks when reviewers lack context; architectural decisions stall when senior engineers are unavailable; knowledge transfer during onboarding can require weeks of pairing and mentorship. Finally, **slow feedback cycles** in both development and review mean that engineers invest significant cognitive effort before learning whether their changes are on the right track.

These bottlenecks are not novel problems; they have long animated discussions about software engineering culture, documentation practices, and organizational structure. However, they are particularly acute in brownfield environments, where legacy constraints and institutional baggage amplify every inefficiency.

## 1.2 Where AI and LLMs Provide Material Leverage

Recent generations of LLMs such as GPT-4, Claude, Gemini, and specialized coding models have demonstrated unexpected capabilities in software engineering tasks. Unlike narrow automation (e.g., IDE autocomplete or syntax highlighting), which addresses surface-level friction, LLMs offer something more powerful: **cognitive augmentation**. They can help engineers externalize and structure knowledge, reason through options, scaffold plans, generate candidate implementations, and provide fast critical feedback before human review.

The key insight is that LLMs excel at tasks that involve restructuring, explaining, planning, and iterating precisely the cognitive work that dominates brownfield engineering. For instance, an LLM given a snapshot of a legacy module, its dependencies, and a feature request can often articulate design options, identify risks, and propose an implementation strategy in minutes, a task that might otherwise require senior engineers to spend hours reading code and whiteboarding. LLMs can draft structured documentation from fragmentary notes, technical tickets, and code samples, helping reduce the time engineers spend on initial drafting. Human review remains essential to ensure correctness and contextual accuracy.

However, the productivity gains from AI are neither automatic nor uniform. A study by Ziegler et al. examining GitHub Copilot's impact on developer productivity found that while developers perceived faster coding, detailed task completion measurements revealed more nuanced results: some tasks saw substantial speedups, while others particularly those requiring significant problem decomposition or domain-specific knowledge showed slower completion times or lower-quality outputs. The variation depended heavily on task type, developer experience, and integration of AI into the workflow. This suggests that **effective AI productivity is not about replacing humans with AI, but rather orchestrating AI tools within a disciplined workflow that preserves human control, accountability, and judgment**.

## 1.3 Introducing the Discover–Define–Deliver (D³) Framework and Dual-Agent Architecture

To systematize how AI should be integrated into brownfield engineering, this paper proposes the **Discover–Define–Deliver (D³) AI Productivity Framework**. The D³ framework presented here is a practitioner-oriented methodology synthesized from existing software engineering best practices and adapted for AI-assisted workflows. This three-phase methodology mirrors how experienced teams naturally approach complex work but augments each phase using a dual-agent architecture: a Builder LLM for creative generation and a Reviewer LLM for critical validation. This separation of concerns mirrors the interaction between a junior engineer and a lead architect. **Crucially, the framework enforces mandatory Human Gates between phases, ensuring that engineers retain final authority over all architectural decisions and code commits.**

The three phases of D³ are:

- **DISCOVER**: building shared understanding and comprehensive context. Engineers and the Builder LLM collaboratively research the codebase, document the current state, identify risks and unknowns, and assemble a shared knowledge artifact. The Reviewer LLM then critiques this research for completeness and clarity, enabling engineers to validate accuracy before planning begins.
- **DEFINE**: designing options, plans, and trade-offs. Using the discovery artifact as foundation, engineers and the Builder LLM propose implementation strategies, cross-repository task plans, testing approaches, and risk mitigations. The Reviewer LLM challenges these plans, identifies hidden coupling, and flags under-specified work, ensuring risks are mitigated before engineers approve the implementation strategy.
- **DELIVER**: executing, reviewing, and validating changes. Engineers write code with Builder LLM assistance, then the Reviewer LLM examines diffs and pull requests for defects and style issues before human review. This AI pre-review streamlines the process, but a human engineer must perform the final review and authorization to merge.

This framework distinguishes **narrow automation** (e.g., "autocomplete the next line of code") from **cognitive augmentation** (e.g., "help me understand what this module does and why it might conflict with my change"). The former provides marginal speedups; the latter can reduce the cognitive load required for complex, open-ended problem-solving by orders of magnitude.

## 1.4 Methodology used for this study

To validate the theoretical challenges of brownfield engineering and assess the efficacy of the proposed framework, I conducted a longitudinal study involving a final cohort of 52 software engineering professionals (Appendix A). The study is exploratory and based on self-reported perceptions rather than controlled task measurements. Participants were recruited through a professional network and voluntarily opted into the study. The respondent cohort represents a highly experienced demographic, with 60% possessing more than 6 years of professional experience and 63% operating within large enterprise environments (3,000+ engineers). Over 84% of respondents reported working primarily in Brownfield or Mixed environments and working on old code bases with tech debt and complexity.

The study was structured in two distinct phases to isolate the impact of the framework:

- **Phase I (Baseline):** The initial survey measured the perceived usefulness and productivity impact of AI tools when used in an "Ad-hoc" capacity, prior to the introduction of any structured methodology.
- **Intervention:** Following Phase I, the full initial cohort was introduced to the D3 AI Productivity Framework and provided with best practices for integrating AI into coding workflows.
- **Phase II (Post-Intervention):** A follow-up survey was conducted 4 weeks after the intervention to measure changes in productivity, cognitive load, and code quality.

While the initial participation pool was larger, the final dataset was filtered to enforce rigorous inclusion criteria. Only respondents who completed both surveys and applied the D³ framework to at least one real-world task were included in the final analysis (N=52), ensuring that the results reflect the actual impact of the methodology rather than theoretical sentiment.

**LLM Tools Used by Participants:**

The study did not mandate specific AI tools, allowing participants to use their preferred LLM platforms. Survey results (Appendix A, Question 4) show that 62% primarily used Anthropic Claude, 19% used OpenAI's ChatGPT, and 19% used Google Gemini Pro. This diversity of tool adoption reflects the current heterogeneous landscape of LLM adoption in enterprise software engineering.

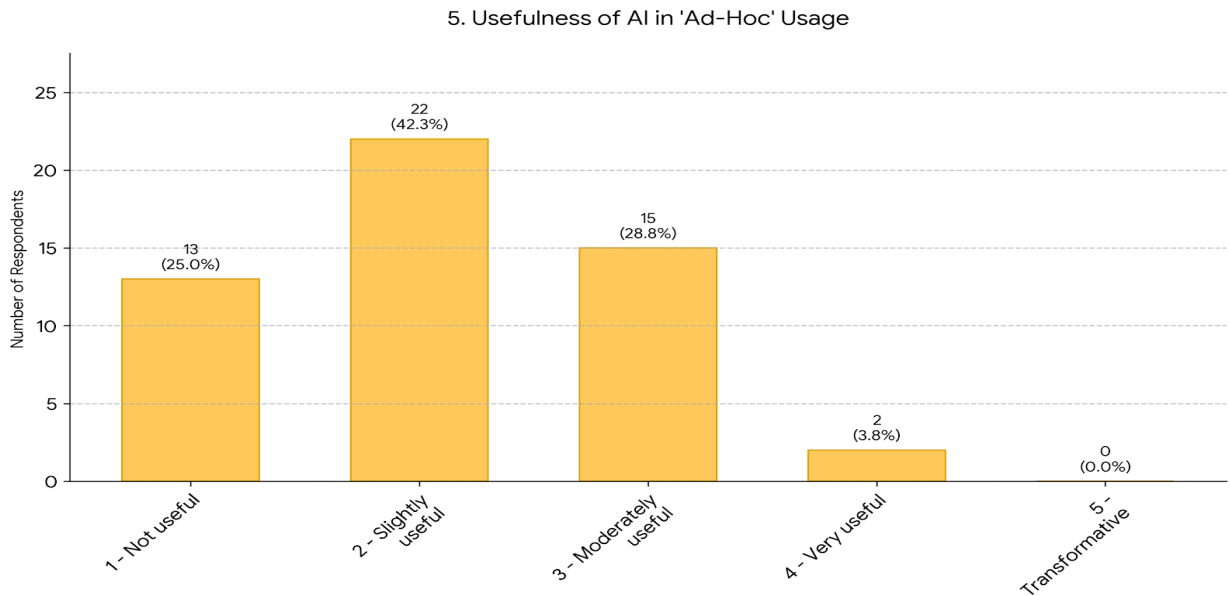

Fig. 1: Survey results highlight that about 68% of respondents did not find AI useful in an ad-hoc

### 1.4.1 Ethics and Data Availability Statement

This study used an anonymous and voluntary survey of adult software engineering practitioners. No personal, identifying, or sensitive information was collected, and the survey platform did not record IP addresses or location data. Participants were informed about the purpose of the study, the voluntary nature of participation, and their right to discontinue at any time prior to submission. Completion of the survey constituted implied consent.

Only aggregated, non-identifiable summary statistics were used in the analysis. Individual raw responses were deleted after aggregation to ensure irreversible anonymity and prevent any possibility of re-identification. Because the study involved only anonymous, minimal-risk survey research, no formal ethics board approval was required under common international research norms.

The aggregated dataset used in this study is fully provided in Appendix A.

# 2. Background and Related Work

## 2.1 Recent Empirical Evidence on AI and Developer Productivity

The past two years have seen a surge of research and industry surveys examining the impact of generative AI on software engineering. A large-scale randomized control trial by Becker et al. (METR) found that experienced developers using early-2025 AI tools on complex open-source tasks actually experienced a **19% slowdown** compared to unassisted developers, largely due to the cognitive load of reviewing and fixing subtle AI errors. This contrasts sharply with the **55% speedup** reported by Ziegler et al. for simpler, isolated tasks, highlighting the 'Brownfield Paradox' where complexity negates raw generation speed.

Complementary research on code quality and correctness has produced mixed signals. Some studies report that AI-assisted code shows comparable defect rates to human-written code, while others note that AI-generated code often requires additional review cycles and has higher rates of subtle logical errors. A 2025 analysis by Harding(GitClear) examining pull requests with AI-generated code found that such PRs required on average 1.3 additional review rounds and 15% more comments from reviewers compared to human-written code, suggesting that velocity gains in writing may be offset by slowdowns in review and validation. These findings underscore the importance of disciplined processes such as the D³ framework that integrate AI into workflow stages where humans retain effective control.

Survey data from Stack Overflow("AI"), and various developer-focused research organizations consistently show that adoption of AI coding assistants is increasing rapidly, with 60–70% of developers at large technology companies having tried AI coding tools. Interestingly, adoption varies significantly by task type: 75–80% of developers report using AI for documentation, explanation, and test generation, while only 40–50% regularly use it for core business logic or security-sensitive code. This pattern suggests that developers intuitively understand where AI adds value versus where human expertise remains critical.

## 2.2 The Role of Documentation in Long-Term Productivity

Documentation is often treated as a secondary concern in engineering organizations a compliance obligation or a nice-to-have that gets deprioritized when schedules slip. However, research on developer productivity paints a different picture. A Google Cloud DORA report found that strong internal documentation practices correlate with higher deployment frequency(Harvey et al.), lower lead times for changes, and lower change failure rates. More specifically, teams that maintain clear design documents, runbooks, and decision records report substantially lower onboarding time for new team members (weeks rather than months) and fewer incidents caused by misunderstanding or misconfiguration. A longitudinal study by Forsgren et el. tracking engineering teams over two years found that investing in documentation infrastructure such as architecture decision records (ADRs), design docs, and API documentation yielded a return on investment within 6–12 months, manifesting as faster bug diagnosis, easier refactoring, and reduced knowledge silos.

The cognitive science underpinning this finding is straightforward: documentation serves as an external store of knowledge, reducing the mental effort required to hold architectural details, rationales, and constraints in working memory. When documentation is weak or absent, engineers must reconstruct this knowledge repeatedly through code reading and informal inquiry, a process that is both cognitively expensive and error-prone. In large organizations with high turnover or distributed teams, this cost compounds.

Yet documentation is also where many large organizations face acute friction. Writing clear, comprehensive documentation requires not just technical knowledge but also communication skill, organization, and the discipline to maintain accuracy over time as systems evolve. Many engineers find documentation writing tedious, a chore that diverts attention from "real" work. As a result, documentation often lags reality or exists only in fragmentary form. This is where AI can provide outsized leverage: by handling the mechanical aspects of writing (structure, coherence, clarity, consistency), AI frees engineers to focus on the substantive work of deciding what should be documented and ensuring accuracy.

## 2.3 Current Approaches to AI in Code Review and Debugging

Code review is a critical bottleneck in large organizations. Reviews serve multiple purposes: catching defects, preserving architectural coherence, distributing knowledge, and providing mentorship. However, reviews are also expensive. A developer waits for a senior engineer's availability, the reviewer must context-switch and reconstruct the mental model of a feature, and if issues are found, multiple rounds of feedback may ensue. Delays in code review directly translate to stalled productivity, reduced parallelism in teams, and frustration.

Recent work on AI-assisted code review shows promise. A 2025 industry case study by Sun et al. reported on automated code review using LLMs at a large technology company, finding that an LLM-based system could identify between 60–75% of the same issues as human reviewers in an automated pre-review pass, catching common defects like incorrect error handling, missing null checks, and violations of coding standards. This reduced the burden on human reviewers, who could then focus on higher-level architectural and design concerns. Similarly, research on debugging with AI shows that LLMs can analyze stack traces, correlate error messages with code, and propose root-cause hypotheses faster than manual debugging, particularly for issues in large codebases where the responsible code may not be obvious.

However, current approaches to AI-assisted review often treat the AI as a standalone tool—a linter replacement that flags issues automatically. The D³ framework proposes a more integrated view: the Reviewer LLM is not meant to replace human review but to preprocess changes, raising critical questions and highlighting risks in a form that human reviewers can quickly assimilate. This preserves the value of human judgment while accelerating the feedback loop.

# 3. Using AI to Write Solid Documents

## 3.1 Why Documentation Pays Off

The case for strong documentation rests on several intertwined observations. First, **documentation reduces onboarding friction and accelerates knowledge transfer**. A new engineer joining a team can, with well-written documentation, reach a productive understanding of core architecture, key business decisions, and common procedures in days rather than weeks. In large organizations, this translates to direct financial impact: if onboarding time drops from 12 weeks to 6 weeks, a team's time-to-productivity increases by a factor of 2, enabling faster project velocity and fewer mistakes during the learning phase. Research shows that structured documentation practices dramatically reduce onboarding time: companies implementing comprehensive onboarding programs reduce time-to-productivity by up to 70% ("The 90-Day"), while Microsoft research found effective onboarding can shorten the typical 3-9 month ramp-up period by 30% ("Onboarding Engineers"). Additionally, a 2024 survey of 50 engineering leaders found that 72% report new hires taking more than 1 month to submit their first 3 meaningful PRs, with those citing longest onboarding times also identifying 'gathering project context' as their top productivity leak ("The 2024 State").

Second, **documentation prevents rediscovery of knowledge**. In the absence of clear documentation, each engineer must individually reconstruct understanding of why a particular design choice was made, what constraints shaped it, and what alternatives were considered and rejected. This is wasteful—the same cognitive work is repeated dozens of times as teams grow or people move between projects. Written decision records and design docs externalize this knowledge, allowing subsequent engineers to learn from prior experience without re-litigating decisions. This is particularly valuable in brownfield environments, where many decisions were made years ago under constraints that may no longer be salient but whose implications persist.

Third, **documentation lowers incident response time and reduces the cost of debugging in production**. When a system behaves unexpectedly, engineers need to understand what the expected behavior should be, what invariants the system maintains, and how different components interact. This understanding is fastest to acquire if documented in runbooks, architecture overviews, and API specifications. Industry research consistently identifies documentation quality as a critical factor in reducing mean time to resolution (MTTR). Up-to-date, easily navigable documentation empowers teams to resolve incidents with greater speed and consistency by minimizing the learning curve and reducing dependency on tribal knowledge ("Complete Guide"). Runbooks that document incident-response procedures are specifically noted as tools that reduce MTTR while simultaneously improving knowledge transfer ("How to Reduce"). A 2024 survey of engineering leaders found that 40% of developers cite 'time to find context' as their top productivity impediment, with 26% of leaders identifying context-gathering as the largest productivity leak across the entire software development value stream ("The 2024 State").

Fourth, **documentation enables distributed teams and async collaboration**. When teams are distributed across geographies or time zones, synchronous knowledge transfer (pairing, in-person discussions) becomes expensive. Documentation allows teams to collaborate asynchronously, with one engineer writing a design doc, other engineers providing feedback asynchronously, and decisions being made and recorded without requiring everyone to be online simultaneously. This is a significant competitive advantage for large, distributed organizations.

Fifth, **documentation supports code quality and reduces technical debt**. When maintainers understand not just what code does but why it was written that way, they are better equipped to make sound refactoring decisions, avoid introducing regressions, and modernize code thoughtfully rather than ad-hoc. The DORA report (Harvey et al.) and associated research (Forsgren et al.) found that teams with strong architectural documentation report lower rates of unintended consequences from refactoring, fewer regressions, and more sustainable pace.

Yet in practice, the blocker to strong documentation is often not lack of understanding but rather cognitive load and articulation skill. Many engineers are technically brilliant but find writing tedious and unnatural. Others understand the value of documentation but struggle with organization and clarity—turning scattered thoughts and bullet points into coherent, accessible prose is harder than it appears. In large organizations, the cumulative effect is pervasive: enormous amounts of technical knowledge and expertise exist in the heads of talented engineers, but relatively little of it is written down in accessible form.

This is precisely where AI offers leverage. By handling the mechanical work of drafting, organizing, and refining text, AI reduces the activation energy required to write documentation. Engineers can work primarily in notes, bullets, and high-level concepts, then use AI to build coherent structure and polished prose. The engineer remains the decision-maker and authority on content, but the tool reduces the friction between thought and articulate documentation.

## 3.2 Where LLMs Help in Documentation

LLMs can play multiple roles in the documentation workflow:

**Structuring and Outlining**: When an engineer has a topic they need to document but feels overwhelmed by the scope, an LLM can suggest a structure or outline. For instance, given a brief description of a feature or system, an LLM can propose sections (Overview, Architecture, Key Decisions, Migration Path, Troubleshooting), hierarchies, and transitions. This gives the engineer a scaffold on which to build, reducing the cognitive load of organization.

**Generating First Drafts**: If an engineer provides raw material—bullet points from a meeting, a set of technical notes, excerpts from code or tickets—an LLM can generate a coherent first draft that incorporates this material. The draft will rarely be perfect or complete, but it can be substantially faster than starting from a blank page. The engineer can then edit, expand, and refine.

**Improving Clarity, Tone, and Coherence**: A second pass with an LLM focused on refining prose can improve readability. An LLM can be asked to simplify technical language, clarify ambiguous passages, make transitions smoother, or adjust tone to match audience expectations. This is particularly valuable for engineers who are competent technical writers but not naturally gifted at prose.

**Enforcing Consistency**: In large documentation systems (such as internal wikis or knowledge bases spanning multiple teams), maintaining consistent terminology, style, and structure across documents is challenging. An LLM can be asked to audit a set of documents for consistency, suggest terminology glossaries, and flag where similar concepts are referred to by different names.

**Generating Supporting Materials**: LLMs can draft summaries, FAQs, quick-start guides, and other derived documentation from comprehensive design docs. This reduces redundant effort—instead of writing both a detailed architecture document and a one-page overview, an engineer can write the detailed version and use an LLM to synthesize the overview.

**Introducing the Reviewer LLM for Documentation**: Beyond generation, a **Reviewer LLM** can strengthen documentation by critiquing drafts. This LLM can be tasked with the perspective of a skeptical, expert reader and asked to identify weaknesses such as:

- Ambiguous or vague statements lacking concrete examples or specifics.
- Missing context that a reader unfamiliar with the system would need.
- Inconsistent terminology or contradictory statements.
- Under-justified decisions or recommendations lacking supporting rationale.
- Assumptions presented as facts.
- Gaps in the narrative where important information is missing.

The Reviewer LLM can also simulate the role of an executive or stakeholder and propose questions they would ask, ensuring the document addresses foreseeable concerns. This is a powerful and low-cost form of peer review, surfacing issues before the document goes to formal review.

**Risks and Mitigations**: However, LLMs in documentation work carry risks that must be actively managed. First, **hallucination and fabrication**: an LLM might confidently assert technical details that are incorrect or invent features or decisions that do not exist. The remedy is that humans must remain the authority on technical content; the LLM is a writing and organization partner, not a source of truth. Second, **outdated recommendations**: if an LLM is trained on general knowledge, it may suggest practices or technologies that are deprecated or inappropriate for the organization. Remedy: provide the LLM with organizational context (coding standards, approved tooling, architecture principles) and ask it to respect those constraints. Third, **leakage of confidential or proprietary information**: if engineers paste sensitive data into an LLM system, that data may be retained or used for training. Remedy: establish clear data policies, use self-hosted or private LLM instances where sensitive work is involved, and redact or anonymize sensitive information in prompts.

## 3.3 Best Practices and Prompting Patterns for Documentation

Effective documentation work with LLMs rests on several principles and patterns that maximize quality and minimize risk.

**Principle 1: Start from Intent, Audience, and Decision Need**. Before writing, be explicit about the purpose. Is this document meant to help a new engineer understand the system? To justify an architectural decision? To provide a runbook for operations? To communicate a business decision? Different intents require different content and structure. Similarly, be explicit about the audience: are they other engineers, product managers, executives, or customers? Is their background deep or shallow? An LLM can tailor content more effectively if given this context.

**Principle 2: Provide Context Rather Than Asking the LLM to Invent**. Instead of asking the LLM to write about a system from scratch (which often results in generic or incorrect content), provide it with raw material: requirements documents, code excerpts, relevant tickets, meeting notes, or existing partial documentation. The LLM then structures and refines this material rather than generating from nothing. This dramatically improves accuracy and relevance.

**Principle 3: Iterate Through Stages Rather Than One-Shot Generation**. Breaking documentation work into stages—outline → rough draft → section refinement → style and clarity pass → Reviewer LLM critique → human final edit—produces better results than attempting to generate a complete, polished document in one prompt. Each stage is focused and benefiting from human refinement before the next stage.

**Prompting Pattern 1: Converting Rough Notes to Structured Documentation**. This pattern helps when an engineer has bullet points or scattered thoughts but needs a coherent document. A Builder LLM prompt might read:

```
I have rough notes for a design document about a payment service refactor.
Here are the notes:
- Current system uses synchronous payment calls, causes latency issues
- Want to move to async with queue
- Need backward compatibility with existing integrations
- Key concern: retry logic and idempotency
- Affects three teams, multiple services depend on payment responses

Create an outline for a design document that covers:
- Problem statement and motivation
- Proposed solution
- Impact on dependent services
- Migration strategy
- Open questions or risks

Make the outline concrete and specific to this project.
```

The LLM responds with a structured outline that the engineer can then flesh out, or can ask the LLM to expand each section.

**Prompting Pattern 2: Turning Meeting Transcripts into Decision Records**. Many critical decisions happen in synchronous meetings, but the outcome is not captured in writing. This pattern extracts the decision and rationale. A Builder LLM prompt:

```
Here is a partial transcript from a meeting where we decided on an approach
to database migration:

[meeting snippet]

Create a concise Architecture Decision Record (ADR) that captures:
1. The decision: what did we decide?
2. Context: what problem did we face?
3. Options considered: what alternatives did we reject and why?
4. Consequences: what are the tradeoffs and downstream impacts?
Keep the ADR to 1-2 pages. Use clear, neutral language.
```

**Prompting Pattern 3: Refining Draft for Clarity and Audience**. After an engineer or Builder LLM produces a draft, a refinement prompt focuses on polish and readability:

```
I have a draft of an internal API documentation page.
The audience is new engineers who may not be familiar with our system.
Here is the draft:

[draft text]

Please:
1. Simplify technical jargon where possible; add a glossary if helpful.
2. Add a concrete example showing how to use the API.
3. Clarify the error handling section--it's vague as written.
4. Check for inconsistent terminology (I noticed we call the main
   object both "Request" and "RequestMessage").
5. Add a "Common Pitfalls" section based on what you infer from the design.

Preserve the core content; just improve clarity and usability.
```

**Prompting Pattern 4: Reviewer LLM Critique for Documentation**. After a draft is ready, a Reviewer LLM examines it with a critical lens:

```
You are an experienced senior engineer and a skeptical reader.
```

```
Here is a design document for a new feature:

[document]

Please critique this document by:
1. Identifying any ambiguous or under-justified claims.
2. Flagging assumptions that should be stated explicitly.
3. Listing questions a reader unfamiliar with this system might have.
4. Suggesting any missing sections or risks that should be addressed.
5. Proposing how an executive or product manager might challenge this
approach.

Be constructive but rigorous. The goal is to strengthen the document
before it goes to formal review.
```

**Implementation and Integration**: In practice, an organization might implement these patterns through a documentation system where engineers have easy access to LLM tools—either through an IDE plugin, a web interface, or a command-line tool. The workflow might look like:

1. Engineer opens a documentation template or blank page.
2. Engineer sketches intent, audience, and rough notes or existing material.
3. Builder LLM generates an outline or first draft (pattern 1 or 2).
4. Engineer reviews and edits the output.
5. Builder LLM refines the draft (pattern 3).
6. Engineer makes final edits.
7. Reviewer LLM examines the final draft, surfacing issues (pattern 4).
8. Engineer addresses Reviewer LLM feedback or decides to accept the current state.
9. **Human Gate: A senior engineer or stakeholder reviews the finalized draft to ensure accuracy and alignment with business goals.**
10. **Upon human approval**, the document is published.

This workflow is designed to be lightweight and non-prescriptive—engineers retain full control and can choose to use as much or as little AI support as suits their preference. The key is providing easy access and clear patterns that reduce friction.

---

# 4. Using AI to Write Code for Brownfield Projects The D[3]

Greenfield projects where engineers start from a blank slate with clear requirements and minimal legacy constraints are ideal for rapid AI-assisted prototyping. However, most large organizations do not live in greenfield. Their work is dominated by brownfield projects: adding features to existing systems, refactoring legacy code, modernizing tech stacks, and maintaining vast, interdependent platforms. These contexts are precisely where AI support is most valuable but also most fragile.

Brownfield work demands that AI tools understand and respect complex, non-obvious constraints: architectural invariants embedded in the codebase, performance characteristics of particular modules, historical decisions that have downstream implications, organizational policies around change management, and the knowledge and availability of domain experts. A generic AI coding assistant trained on public GitHub repositories may generate syntactically correct code that violates these constraints or introduces subtle coupling. The D³ framework addresses this by anchoring AI work in detailed, localized context and making human review and decision-making mandatory at key gates.

## 4.1 The D³ Methodology for AI-Assisted Coding in Brownfield Projects

The **Discover–Define–Deliver (D³) AI Productivity Framework** represents a disciplined approach to applying AI in complex brownfield environments, where understanding existing systems and managing risk are as important as writing new code. The framework is grounded in how experienced teams naturally approach large refactors or feature additions but systematically applies AI and LLM support to each phase.

**To execute this systematically, the framework relies on a Dual-Agent Architecture that separates concerns between two distinct AI roles:**

- **Builder LLM:** tasked with generation, exploration, and drafting. This role takes prompts from humans and produces research documents, implementation plans, code snippets, documentation, and other creative outputs. The Builder LLM is optimized for coverage, breadth, and rapid hypothesis generation.
- **Reviewer LLM:** tasked with critique, validation, and stress-testing. This role examines artifacts produced by humans or the Builder LLM, identifies weaknesses, raises critical questions, and proposes refinements. The Reviewer LLM functions as a principal engineer or skeptical peer, bringing a different perspective and catching oversights.

This dual-agent approach creates discipline across all three phases. The Builder LLM generates candidates and explores options; the Reviewer LLM stress-tests these outputs and raises critical questions. Together, they create a workflow that preserves human judgment while accelerating task completion

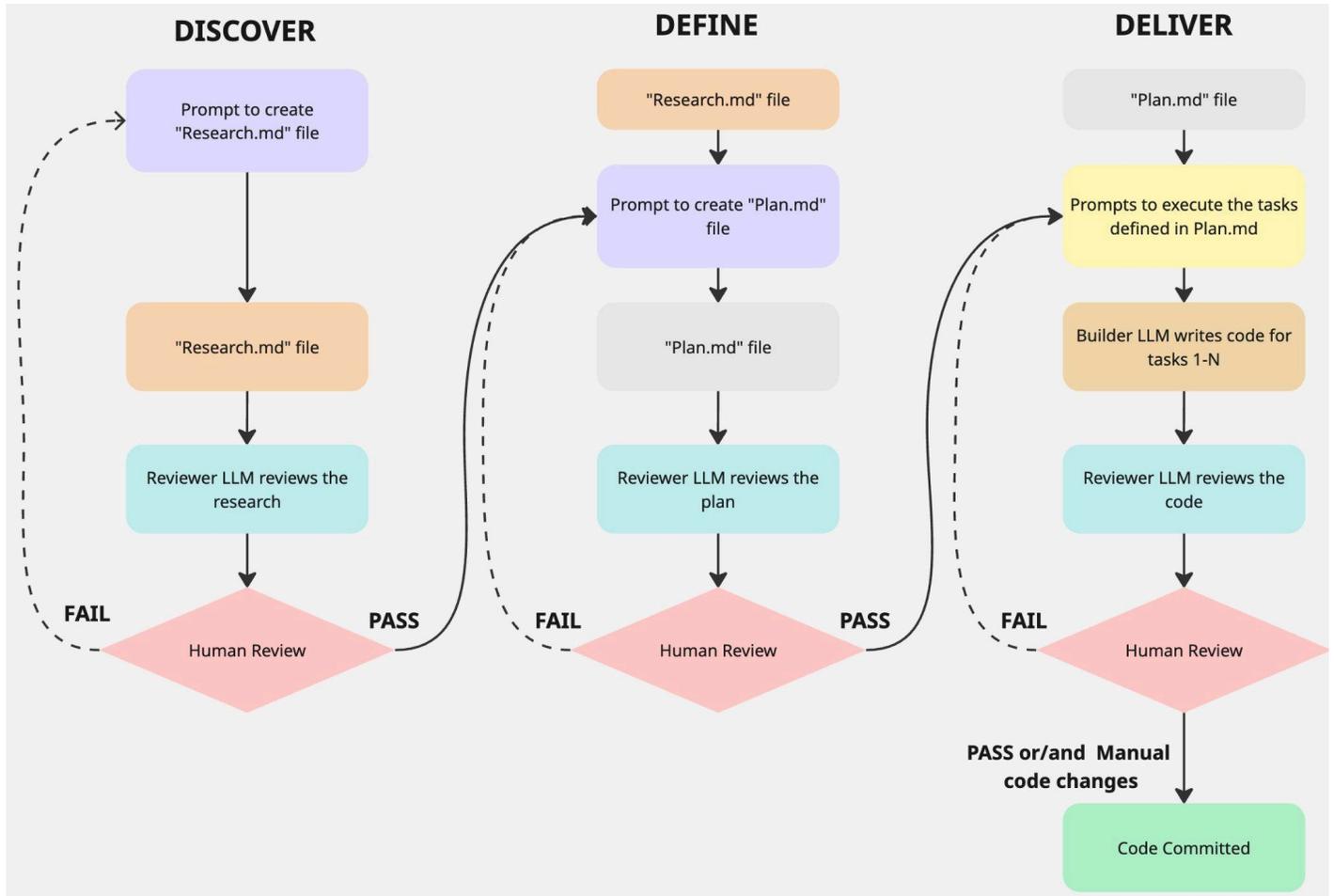

Fig. 2: Flow Diagram of D3 Framework

**4.1.1 DISCOVER – Repository Research and Context Building**

The **DISCOVER** phase begins with the engineer and the Builder LLM collaboratively exploring the codebase. The goal here is to build a **"research file"**, which is not the final design but an artifact used by engineers to gain productive insights about the code base. The research file serves as the foundation for all subsequent work. It captures the current state of the system, architectural constraints, dependencies, and known risks. This artifact becomes the shared understanding that both AI agents and human engineers reference throughout the project. efficient.

**Step 1: Create Research File with Comprehensive Prompt**

Begin by creating a `RESEARCH.md` file using a single comprehensive prompt to your Builder LLM:

```
You are a senior software architect conducting comprehensive brownfield
```

```
codebase research.

Your task is to create a complete RESEARCH.md file that covers all aspects
of the current system state.

ANALYSIS REQUIREMENTS:

1. REPOSITORY IDENTIFICATION & SCOPE
   - Identify all repositories that handle or interact with [the system
being changed]
     - For each repository, document:
       * Primary responsibilities
       * Relationship to [the system]
       * Owner/team
       * Current version/tech stack

2. DEPENDENCY MAPPING
   - Incoming dependencies (what services/modules call this system?)
   - Outgoing dependencies (what does this system call?)
   - Shared libraries and their versions
   - Database schemas and tables accessed
   - External system integrations
   - Trace the call graph through at least 3 levels of dependencies
   - Identify transitive dependencies that might be affected

3. CURRENT SYSTEM ARCHITECTURE
   - Document the current data models (all fields, types, relationships)
   - Document all API endpoints (paths, methods, parameters, responses)
   - Document database schemas related to [the system]
   - Document cache layers and their purpose
   - Include code examples from the actual codebase

4. ARCHITECTURAL DECISION EXTRACTION
   Review the codebase and extract:
   - Any ADR (Architecture Decision Record) files
   - Code comments explaining "why" decisions were made (not just "what")
   - Git commit messages mentioning [relevant keywords]
   - Ticket references in code comments (e.g., JIRA-1234)
   - TODO or FIXME comments related to [the system]
   - For each finding, document: Decision Made | Rationale | Date (if
available) | Current Status

5. PERFORMANCE & OPERATIONAL CHARACTERISTICS
```

```
        - Current performance metrics (latency, throughput if documented)
        - Scalability characteristics
        - Known bottlenecks or limitations
        - Monitoring and alerting in place
        - Incident history (if documented in code/comments)

6.  RISK IDENTIFICATION
    Identify and categorize:
        - KNOWN RISKS: Clearly documented concerns or limitations
        - UNKNOWNS: Areas with insufficient documentation or unclear behavior
        - ASSUMPTIONS: Things the current system assumes that might not hold
        - COMPLIANCE/SECURITY: Any regulatory or security requirements mentioned
        - PERFORMANCE RISKS: Potential performance implications of changes

7.  AFFECTED TEAMS & STAKEHOLDERS
        - Which teams own affected repositories?
        - Which teams depend on this system?
        - Are there external stakeholders or customers affected?

8.  DATA FLOW & INTEGRATION POINTS
        - Trace how data flows through the system
        - Document integration points with other services
        - Identify synchronous vs asynchronous interactions
        - Document error handling patterns
```

**Step 2: Reviewer LLM Critique**

Once a draft research document is assembled, the Reviewer LLM examines it for completeness and clarity. Questions the Reviewer LLM might raise: "You identified three services that depend on Sessions, but did you check for transitive dependencies? Are there any data pipelines or scheduled jobs that might be affected? Did you consider how this change affects local development and testing?"

**Step 3: Human Gate**

Before proceeding to the DEFINE phase, a human engineer must review and approve the research document to validate that the Builder LLM's context is accurate, the Reviewer LLM's concerns have been addressed, and all other factors which might be missed by LLMs are taken care of.

**Survey results about impact of DISCOVER phase:**

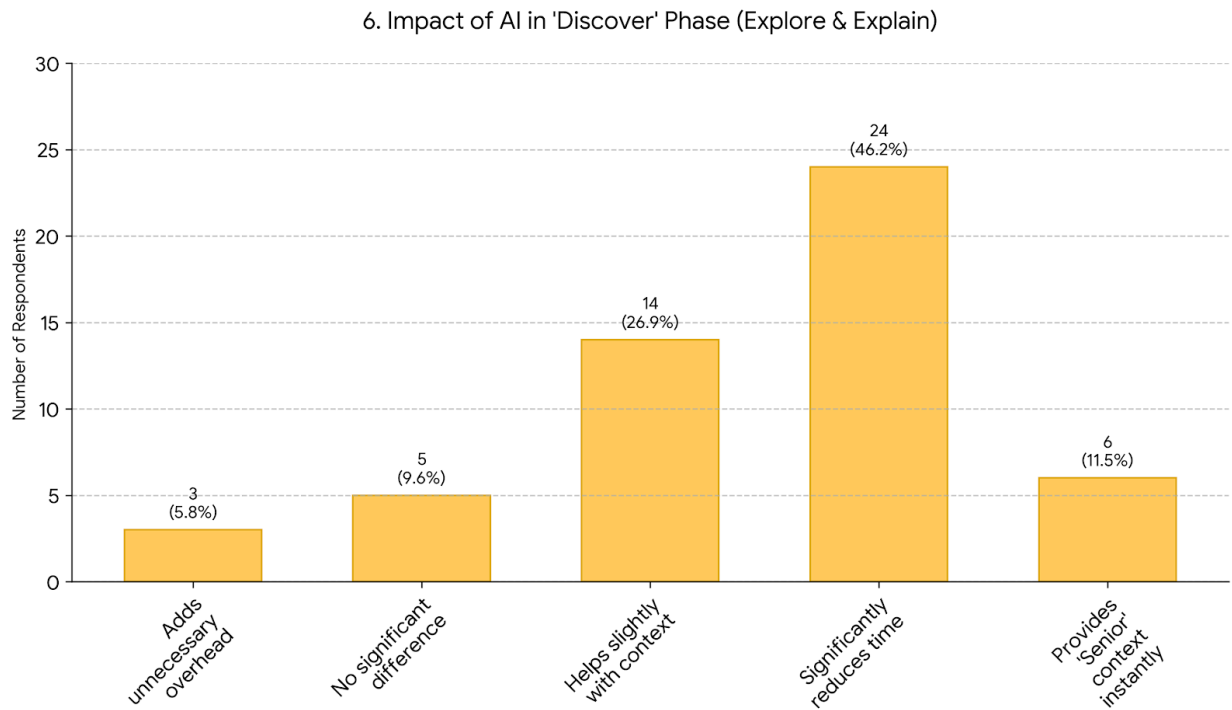

Fig. 3: Survey results highlight that about 57% of respondents found Discover phase useful to understand legacy code, their time spent was reduced and some even find senior engineer level context.

### 4.1.2 DEFINE – Planning, Design, and Risk Assessment

With the context established in DISCOVER, The output of DEFINE is a comprehensive implementation plan that an engineering team can execute with high confidence, knowing the key decisions have been made, risks have been identified, and the work can be parallelized across teams.

**Step 1: Create Plan File with Comprehensive Prompt**

Create a `PLAN.md` file and use the Builder LLM for a specific task:

```
You are a senior software architect designing an implementation strategy.

CONTEXT:
- Research document: [Attach or reference RESEARCH.md]
- Project goal: [Describe the change]
- Constraints: [Timeline, resources, risk tolerance]

Your task is to create a comprehensive PLAN.md file.
```

```
DESIGN EXPLORATION:

Propose 3 distinct approaches for [the change]. For each approach, detail:

1. IMPLEMENTATION STRATEGY
   - High-level approach and methodology
   - Key architectural decisions
   - Technology choices and rationale

2. MIGRATION PATH
   - Phased rollout strategy
   - Backward compatibility approach
   - Coexistence period (if applicable)

3. REQUIRED CODE CHANGES
   - Describe what changes are needed
   - Estimate complexity (Small/Medium/Large)

4. TESTING APPROACH
   - Unit testing strategy
   - Integration testing requirements
   - E2E testing scenarios
   - Performance testing needs

5. PROS AND CONS
   - Benefits of this approach
   - Drawbacks and limitations
   - Risk assessment

6. EFFORT ESTIMATE
   - Engineering days per repository
   - Total team effort
   - Timeline estimate

7. RISK LEVEL
   - Overall risk: [Low/Medium/High]
   - Key risk factors
   - Mitigation strategies

After presenting all 3 approaches, recommend one with clear justification based on:
- Risk vs. benefit tradeoff
```

```
- Alignment with constraints
- Technical feasibility
- Team capability

Be specific to our tech stack and constraints documented in RESEARCH.md.

## Based on your recommended approach, create detailed task breakdowns.

TASK 1: [Clear, actionable title]
- DESCRIPTION: [What needs to be done - be specific]
- ACCEPTANCE CRITERIA:
  * [Criteria 1]
  * [Criteria 2]
  * [Criteria 3]
- FILES LIKELY TO CHANGE:
  * [path/to/file1.ext] - [what changes]
  * [path/to/file2.ext] - [what changes]
- DEPENDENCIES: [Other tasks that must complete first]
- TESTING REQUIREMENTS:
  * Unit tests: [specific tests needed]
  * Integration tests: [scenarios to cover]
  * Performance tests: [benchmarks to hit]
- INTEGRATION POINTS: [How this connects to other tasks]
- ESTIMATED EFFORT: [X hours/days]
- RISK LEVEL: [Low/Medium/High]
- ROLLBACK PLAN: [How to undo if needed]

TASK 2: [Next task...]

TASK 3: [Next task...]
```

**Step 2: Reviewer LLM Critique**

You can use below prompt to critique the `PLAN.md` created:

```
You are a staff engineer reviewing this implementation plan before work begins.
Your role is to challenge the plan and identify potential failure modes.

Review the PLAN.md and provide critical feedback:

1. HIDDEN COUPLING
```

```
       - Are there dependencies between tasks not explicitly captured?
       - Could changes in one repository break another in non-obvious ways?
       - Are there shared resources (databases, caches, queues) with potential
conflicts?

2. TESTING GAPS
       - Are there edge cases not covered by the testing strategy?
       - Have we tested failure scenarios (network partitions, service
outages)?
       - Is the test data representative of production?
       - Can we detect performance regressions?

3. SINGLE POINTS OF FAILURE
       - Are there steps where failure would block the entire project?
       - Do we have contingency plans?
       - Are critical path tasks assigned to available people?

4. ROLLBACK SAFETY
       - Can we truly roll back at each phase?
       - What about data consistency during rollback?
       - Have we tested rollback procedures?
       - Are there one-way doors (irreversible changes)?

5. SECURITY & COMPLIANCE
       - Have we considered security implications?
       - Are there compliance requirements (audit logging, data retention)?
       - Have security and compliance teams reviewed?

For each concern, provide:
- SEVERITY: [Critical | High | Medium | Low]
- ISSUE: [Clear description of the problem]
- IMPACT: [What could go wrong if not addressed]
- RECOMMENDATION: [Specific changes to the plan]
- BLOCKER: [Yes/No - does this prevent proceeding?]

Be constructively critical. The goal is to strengthen the plan before
execution.
```

**Step 3: Human Gate**

The implementation plan must be formally reviewed and approved by a human engineer. This ensures the Builder LLM's proposed strategy is sound, the solution is realistic, and the risks mitigation is adequate.

**Additional Considerations:**

To check the impact on downstream repositories, additional impact analysis can be. As part of this process a cross-repository context can be given to the Builder LLM. All of those repositories should have their own `research.md` file which can be used by Builder LLM to create an artifact called `impact.md` for each repository. Furthermore, based on all impact files, the Builder LLM can be used to improve the `plan.md` file.

**Survey results about impact of DEFINE phase:**

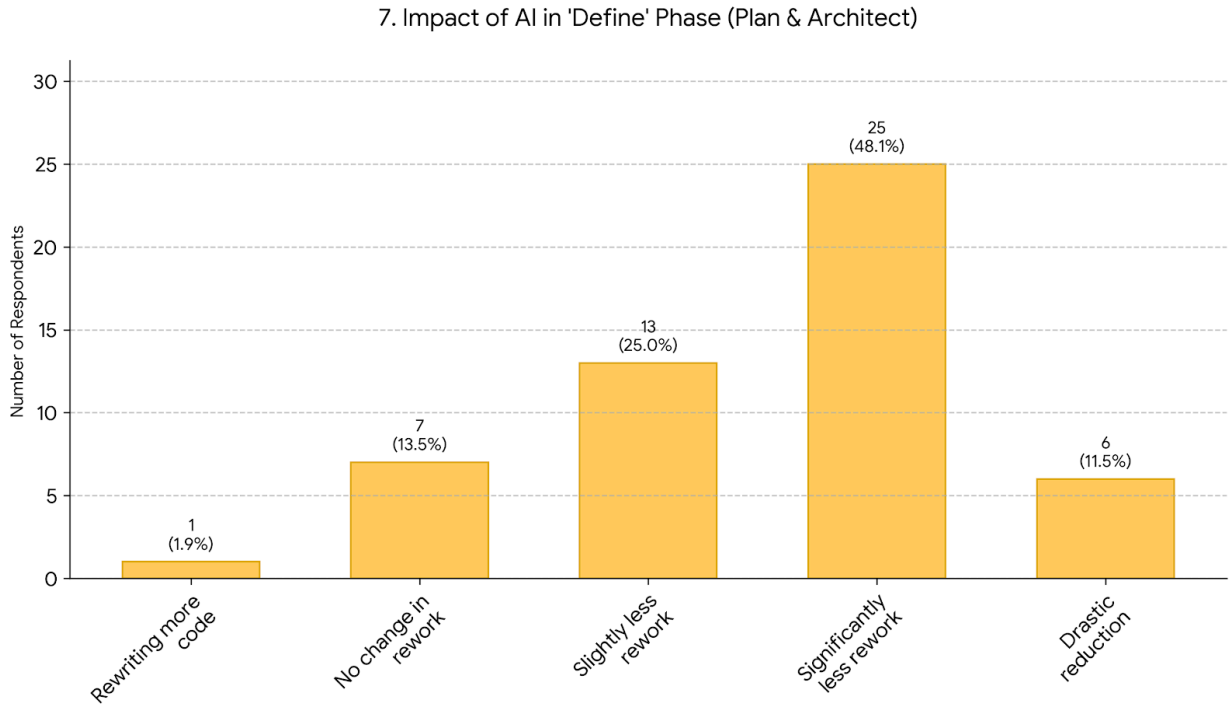

Fig. 4: Survey results highlight that the Define phase reduced the re-work required while using LLMs in coding for about 83% of respondents.

The Reviewer LLM helps de-risk the plan before execution, surfacing gaps that might not be obvious to a single engineer.

### 4.1.3 DELIVER – Execution, Code Review, and Validation

With the plan in place, the **DELIVER** phase focuses on turning the plan into reality: writing code, testing, reviewing, and validating.

The DELIVER phase proceeds as follows:

1. **Task Execution with Builder LLM Support**: For each task from the plan, the engineer uses the Builder LLM to generate code. For instance, given a description of "Implement new Token API endpoint in Auth Service that issues JWT tokens," the Builder LLM can generate skeleton implementations, handle common patterns, and provide candidate code that the engineer reviews and refines. The engineer retains full control—they read, edit, and verify the code before committing.
2. **Pre-Review with Reviewer LLM**: Once code is drafted, the engineer can submit diffs to the Reviewer LLM before human code review. The Reviewer LLM examines the changes and reports:
   - Summary of what changed and why.
   - Potential defects (off-by-one errors, missing null checks, race conditions).
   - Style or architectural violations.
   - Missing tests or coverage gaps.
   - Suggestions for improvement or alternative approaches.
3. This pre-review catches many issues before human review, reducing the burden on reviewers.
4. **Human Review and Judgment**: Senior engineers review the code, focusing on architectural soundness, security implications, and adherence to organizational standards. The human reviewer provides the final approval required to merge the code, serving as the ultimate check against both human and AI errors.
5. **Iterative Refinement**: If issues are found, the engineer addresses them, potentially with Builder LLM assistance for generating alternative implementations, and resubmits.
6. **Testing and Validation**: As code lands, continuous integration tests run automatically. The plan from DEFINE specifies integration tests, canary deployments, and monitoring. If issues arise, they are caught quickly, and the team can roll back or iterate.

The output of DELIVER is high-quality, well-tested code that has been reviewed by both AI and humans and validated against the plan.

### 4.1.4 How D³ Reduces Cognitive Load and Scales Execution

The D³ methodology achieves several critical goals:

**Reduces the cost of understanding legacy code**: By systematically exploring the codebase in DISCOVER, the team builds shared understanding up front, reducing the need for individual engineers to repeatedly re-read and re-understand the same systems. This is particularly valuable for large teams or for engineers joining mid-project.

**Makes work parallelizable and delegable**: By decomposing work into clear tasks in DEFINE, senior engineers can delegate tasks to junior engineers with confidence. A junior engineer with a clear task description and a research document can execute effectively. They are not left to figure out scope or risk on their own.

**Preserves human control and accountability**: At each stage, humans make the key decisions. The Builder LLM generates candidates; humans choose. The Reviewer LLM raises concerns; which helps a human reviewer to be more efficient in finding issues.. AI is a tool that augments human capability, not a replacement for human judgment.

**Captures productivity gains while managing quality**: By offloading cognitive work (exploration, planning, code generation, and review support) to AI, while keeping humans in control of decisions, the D³ approach achieves productivity gains in some contexts. These results are highly task-dependent and should not be generalized without controlled evaluation.

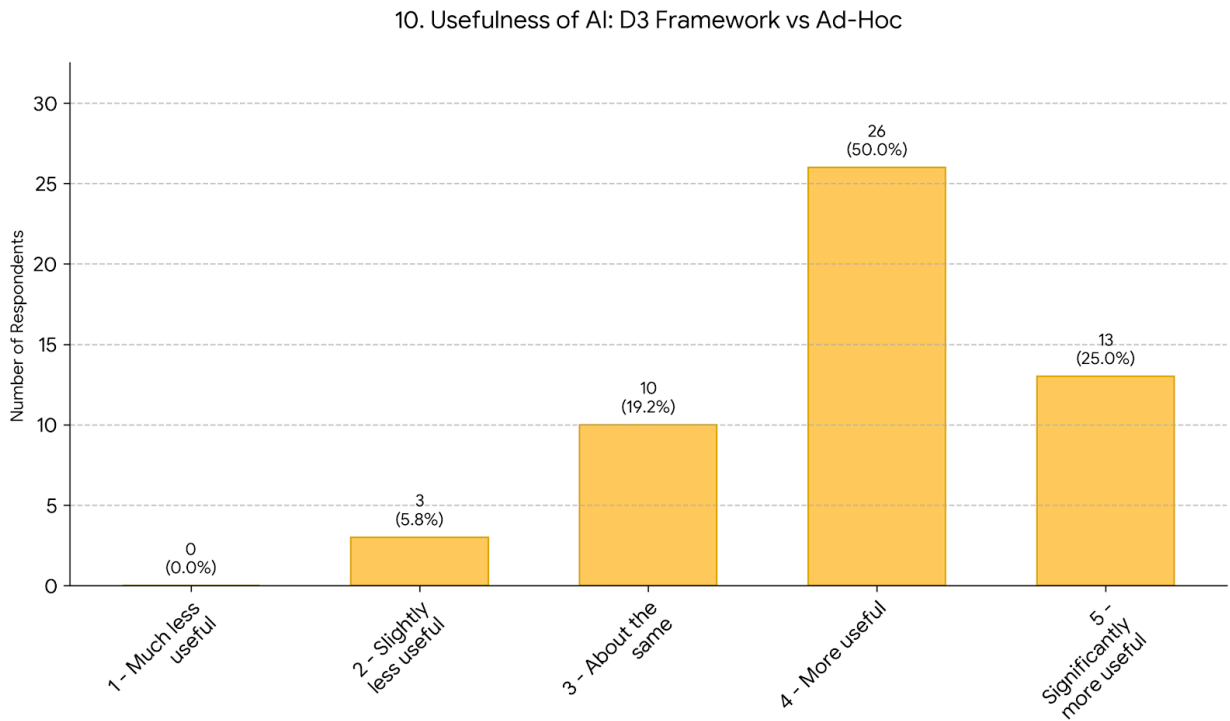

Fig. 5: Survey results demonstrate that 75% of the respondents found the "D3 framework for AI productivity" more useful than using LLMs for coding on adhoc basis..

## 4.2 Prompting Techniques and Multi-Context Usage

Effective use of AI in coding requires understanding a range of prompting techniques that structure how the LLM thinks and responds. These techniques have emerged from both research and practitioner experience and can be adapted to brownfield coding tasks.

### 4.2.1 Core Prompting Techniques

**Meta-Prompting**: Explicitly tell the model how to approach the problem, what role to play, what constraints to respect, and what failure modes to avoid. For instance:

```
You are an expert software architect reviewing a legacy codebase.
Your goal is to understand the impact of adding a new database column to
the User table.

When you respond:
- Be precise and specific to this codebase; avoid generic advice.
- Flag assumptions where you lack information.
- Prioritize identifying hidden dependencies and risks.
- Do not recommend large refactors; work within existing architecture.
- If you are uncertain about a detail, say so explicitly.

Here is context about the User table and its usage...
```

This meta-prompt shapes the LLM's perspective and constraints before it tackles the substantive problem.

**Prompt Chaining and Recursive Prompting**: Break a complex problem into a sequence of steps, where each step's output feeds into the next. For instance, to design a refactor of a large module:

1. First prompt: "Describe the current structure and responsibilities of this module."
2. Second prompt: "Given this structure, what are the main pain points or design issues?"
3. Third prompt: "Given the issues, propose two refactoring approaches and their tradeoffs."
4. Fourth prompt: "Given the two approaches, select the one that best balances simplicity, risk, and benefit. Justify your choice."
5. Fifth prompt: "Given the selected approach, break it into implementable tasks."

Each prompt builds on the previous response, iteratively refining the solution.

**One-Shot and Few-Shot Examples**: Show the LLM examples of the style, structure, or level of detail you want in the response. For instance:

```
Here are examples of good task descriptions for code changes:

EXAMPLE 1:
Task: Refactor UserService.authenticateUser to use TokenValidator
What: Replace direct token validation logic with call to new
TokenValidator.validate().
Why: TokenValidator will handle multiple token types (JWT, OAuth, Session)
consistently.
Files affected: UserService.java, TokenValidator.java
Tests needed: Unit test for UserService.authenticateUser with mocked
```

```
TokenValidator.
Estimated effort: 2 hours.

EXAMPLE 2:
Task: Add audit logging to PaymentService
What: Wrap PaymentService.processPayment with logging that records amount,
status, timestamp.
Why: Audit trail required for compliance and troubleshooting payment
issues.
Files affected: PaymentService.java, AuditLogger.java
Tests needed: Integration test verifying audit log entry created for each
payment.
Estimated effort: 3 hours.

Now, generate task descriptions for the Auth refactor plan with similar
detail and structure.
```

By showing examples, you guide the LLM toward the structure and detail level you need.

**System-Level Instructions and Feedback Loops**: If you are using an LLM repeatedly (e.g., regularly asking for code reviews or design feedback), you can establish standing instructions that apply across interactions:

```
System instructions for this conversation:
1. When I ask for code generation, assume the language is Python 3.10+.
2. Prefer type hints and modern idioms over backward compatibility.
3. If you generate code, include docstrings for all functions and classes.
4. When reviewing code, focus on clarity and maintainability, not
micro-optimizations.
5. Flag security concerns prominently. Error handling should be explicit.
6. Assume the reader is familiar with this codebase; do not explain basic
patterns.
```

These standing instructions reduce the need to repeat context in every prompt.

**Determinism vs. Randomness**: LLMs have a "temperature" parameter that controls randomness. For predictable, consistent responses (e.g., code generation, data extraction), use lower temperature (e.g., 0.2–0.3). For exploratory or creative tasks (e.g., brainstorming design options), use higher temperature (e.g., 0.7–0.9) to encourage variety. Be explicit about this when using LLM systems:

```
Generate three alternative approaches to caching the User object.
```

```
Temperature: 0.8 (encourage creative variety)

For each approach, then generate the most likely implementation.
Temperature: 0.2 (consistent, predictable code)
```

4.2.2. Multi-Context Prompting: Images and Diagrams

Recent LLMs have become multimodal, capable of accepting not just text but also images, diagrams, and sometimes other modalities. This opens new possibilities for brownfield engineering, where legacy systems are often best understood visually.

**Architecture Diagrams and Code Understanding**: An engineer can provide an architecture diagram—showing services, databases, APIs, and data flows—along with a text description, and ask the LLM to explain interactions, identify bottlenecks, or propose changes. For instance:

```
Here is an architecture diagram of our user authentication system:
[image: diagram showing Auth Service, User Service, API Gateway, Session Cache, Database]

I want to migrate from Session-based to Token-based authentication.
Based on this diagram, what are the critical paths I need to change,
and what services could be affected by changes to the Auth Service?
```

The LLM can now reason about the visual structure alongside textual context, often providing more accurate and relevant responses than text alone.

**UI Mockups and Frontend/Backend Integration**: Product designers create mockups; engineers need to understand what backend changes are required. An engineer can provide a mockup of a new UI along with a description of the backend service, and ask the LLM to propose backend changes:

```
Here is a mockup of a new user preferences page:
[image: mockup showing new fields for notification frequency, timezone, etc.]

The backend User service currently returns a UserProfile with name, email, age.
What changes to the UserProfile schema and API would be needed to support this UI?
What migration strategy would you propose?
```

**Log and Error Screenshots**: When debugging a complex issue, engineers can provide screenshots of logs, stack traces, or error messages alongside code snippets, and ask the LLM to diagnose the issue:

```
I'm seeing this error in production:
[image: screenshot of log lines showing "NullPointerException at
UserService.getUserPermissions:42"]

Here is the relevant code:
[code snippet]

What is likely causing this error, and what would you recommend as a fix?
```

The LLM can often spot issues more quickly with visual context than with text alone.

**Database Schema Diagrams**: For brownfield systems, understanding the database schema and relationships is critical. An engineer can provide an ER diagram or a text schema definition alongside a feature description, and ask the LLM to propose schema changes:

```
Here is the current database schema:
[ER diagram or text schema showing User, Account, Transaction, Audit tables
and relationships]

We want to add support for multi-currency transactions.
What schema changes would you recommend, and what migration strategy?
```

### 4.2.3. Voice-Style Prompting and Conversational Workflows

Some modern LLM platforms support voice input, and engineers can speak their thoughts rather than typing. This is particularly valuable for exploratory problem-solving where the engineer is thinking out loud.

A voice interaction might look like this (transcribed):

```
ENGINEER: Okay, I need to understand why this payment processing is slow.
Let me talk through what I know. We have a PaymentService that takes a
request,
validates the card, calls the payment gateway, logs the result, and
returns.
In production, end-to-end latency is 800ms, but I'd expect it to be around
200ms.
```

```
ASSISTANT: That's a significant gap. Do you know where the bottleneck is?
Have you profiled the service?

ENGINEER: Not yet. Let me describe the flow more carefully.
[describes flow in detail]

ASSISTANT: Based on what you've described, there are several potential
bottlenecks:
card validation might be synchronous and slow, the payment gateway might
have
high latency, or logging might be blocking. What metrics or logs could help
us narrow this down?

ENGINEER: Good point. I think we have logs in Cloud Logging.
Can you help me figure out what queries to run?

ASSISTANT: Yes, here are some queries to isolate which step is slow...
```

This conversational, exploratory mode is particularly valuable for complex debugging or architectural discussions, where iterative back-and-forth is more productive than a single long prompt.

## 4.3 Recommended Use Cases for AI in Engineering

Not all coding tasks benefit equally from AI assistance. Understanding which tasks are high-value candidates, and where AI actually adds productivity, is essential to implementing D³ effectively. The use cases in this section are informed by best practices documented in industry guide("Guide to AI").

### 4.3.1 Stack Trace Analysis and Debugging

**The Use Case:** A developer encounters a production error—a stack trace, error logs, or a user report of unexpected behavior. They need to understand the root cause and propose a fix.

**Where It Fits in D³:** DELIVER phase (debugging incident responses) and DISCOVER phase (understanding unfamiliar code).

**AI Roles:**

**Builder LLM:** Analyzes the stack trace, correlates it with code, and proposes likely root causes. For instance, given a NullPointerException and the relevant code, the LLM can identify that a field is not being initialized under certain conditions.

**Reviewer LLM:** Examines proposed fixes for correctness and completeness. "You propose adding a null check, but did you consider how this affects callers? Are there other code paths that could fail in the same way?"

**Why It Works:** Stack traces contain concrete information (function names, line numbers, exception types). LLMs are good at correlating this information with code and identifying likely causes.

**Pitfalls to Avoid:** Debugging with AI can sometimes lead to premature conclusions. An LLM might suggest the first plausible cause without exploring alternatives. Mitigate by asking the LLM to propose multiple hypotheses and explain how to test each.

### 4.3.2 Refactoring and Legacy Code Modernization

**The Use Case:** An engineer needs to refactor a legacy module—improve structure, reduce complexity, migrate to a newer framework or pattern—while maintaining behavior and minimizing risk.

**Where It Fits in D³:** DISCOVER phase (understanding the legacy code), DERIVE phase (planning the refactor), DELIVER phase (implementing changes).

**AI Roles:**

**Builder LLM:** Analyzes the existing code, identifies improvement opportunities, and generates candidate refactored versions. For instance, given a large function with nested conditionals, the LLM can extract helper functions, simplify logic, and propose more readable structure.

**Reviewer LLM:** Compares the original and refactored code, verifying that behavior is preserved and that the refactored version is genuinely simpler or better. "The refactored version reduces lines of code by 30%, but I notice you changed how errors are handled. Is this intentional? Does the new approach still log errors properly?"

**Why It Works:** Refactoring benefits from understanding both the old and new code and reasoning about equivalence. LLMs are capable of this.

**Pitfalls to Avoid:** Refactoring is risky; even small bugs can cause regressions. Mitigate by requiring comprehensive tests before and after refactoring, and by using the Reviewer LLM to explicitly verify that behavior is preserved.

### 4.3.3 Mid-Loop Code Generation

**The Use Case:** An engineer has a partially written function or code outline and needs the AI to complete scoped blocks of code without rewriting from scratch.

**Where It Fits in D³:** DELIVER phase (implementing tasks and filling in code gaps).

**AI Roles:**

**Builder LLM:** Receives a code outline or function signature with description, and generates the implementation. Given context and constraints, the LLM fills in logic, handle edge cases, and completes the function body in a way consistent with the outline.

**Reviewer LLM:** Examines the generated implementation for logical correctness, efficiency, and alignment with the outline intent. "You completed the function, but I notice you're not handling the null case for the optional parameter. Should you add that?"

**Why It Works:** Mid-loop generation is highly efficient because the engineer has already made key decisions (function signature, algorithm approach, integration points) and only needs the LLM to fill in the details. This combines the structure and judgment of the engineer with the LLM's speed at coding. Research shows this approach reduces coding time by 30–50% while maintaining high quality because the overall shape is human-determined.

**Pitfalls to Avoid:** The generated code might not match the engineer's intended algorithm or might introduce inefficiencies. Mitigate by providing clear comments and constraints in the outline, and by having the engineer review the output before committing.

### 4.3.4 Test Case Generation

**The Use Case:** An engineer has written code or is reviewing code and needs to ensure it is tested. Rather than manually writing test cases, they ask an LLM to generate tests.

**Where It Fits in D³:** DELIVER phase (generating tests for new code), DISCOVER phase (adding tests to legacy code lacking coverage).

**AI Roles:**

**Builder LLM:** Generates unit, integration, or property-based tests from code. Given a function signature and description, it can generate test cases for normal paths, edge cases, and error conditions.

**Reviewer LLM:** Examines the generated tests for coverage gaps, weak assertions, or testing anti-patterns. "You generated tests for the happy path, but I notice you didn't test what happens if the input is null or the external service is down. Should you add those?"

**Why It Works:** Test generation is largely mechanical—identify input scenarios, assert expected outputs—and LLMs excel at this.

**Pitfalls to Avoid:** Generated tests can have low-quality assertions or miss important edge cases. Mitigate by having humans review and edit generated tests, and by measuring test coverage and mutation scores to identify gaps.

4.3.5 Learning New Techniques and Frameworks

**The Use Case:** An engineer encounters an unfamiliar framework, library, language pattern, or piece of legacy code and needs to understand how it works quickly.

**Where It Fits in D³:** DISCOVER phase (understanding existing systems), DELIVER phase (learning a new framework for implementation).

**AI Roles:**

**Builder LLM:** Explains code, translates between frameworks (e.g., "here is equivalent code using the new framework"), and generates concrete examples. Acts as a tutor or reference guide.

**Reviewer LLM:** Verifies that explanations are correct and complete. "Your explanation of how this dependency injection framework works is clear, but I notice you didn't mention the initialization order—is that important?"

**Why It Works:** Explaining code is a natural task for LLMs, and having an interactive partner to ask questions to (rather than reading documentation) accelerates learning. This is particularly valuable for onboarding, for engineers working on unfamiliar systems, and for accelerating adoption of new technologies or patterns within an organization.

**Pitfalls to Avoid:** Explanations can be incorrect or misleading, especially for obscure frameworks or bleeding-edge technologies. Mitigate by asking the LLM to provide concrete examples and by verifying explanations against actual behavior in the codebase.

4.3.6 Complex Query Writing (SQL, Regex, CLI Commands)

**The Use Case:** An engineer needs to write complex database queries, regular expressions, command-line commands, or other domain-specific languages but finds them tedious or error-prone to construct manually.

**Where It Fits in D³:** DELIVER phase (implementing tasks that require complex queries or commands).

**AI Roles:**

**Builder LLM:** Generates complex SQL queries, regex patterns, CLI commands, or similar based on natural language descriptions. For instance: "Write a SQL query that finds all users who joined in the last 30 days, have made at least 5 purchases, and have not unsubscribed from emails."

**Reviewer LLM:** Examines generated queries for correctness, efficiency, edge cases, and security implications. "The query you generated will work, but it will perform a full table scan on the users table. For large tables, this could be slow. Would adding an index on the join date improve performance?"

**Why It Works:** Complex domain-specific languages have high cognitive load and are error-prone to construct manually. LLMs are trained on large amounts of SQL, regex, and command-line examples and can generate correct, idiomatic queries quickly.

**Pitfalls to Avoid:** Generated queries might be inefficient, incorrect, or have security issues (e.g., SQL injection vulnerabilities if not carefully constructed). Mitigate by having the engineer test queries before deployment and by using the Reviewer LLM to highlight performance and security concerns.

### 4.3.7 Code Documentation Generation

**The Use Case:** Code lacks comments or documentation. Rather than manually writing comments, an engineer asks an LLM to generate them.

**Where It Fits in D³:** DELIVER phase (documenting new code), DISCOVER phase (adding documentation to legacy code).

**AI Roles:**

**Builder LLM:** Generates comments, docstrings, and high-level documentation from code. Produces clear descriptions of function purpose, parameter meanings, return values, exceptions raised, and usage examples.

**Reviewer LLM:** Examines generated documentation for accuracy and usefulness. "The docstring for this function is clear, but it doesn't mention the performance implications of the algorithm—should it? Also, are there any edge cases that should be documented?"

**Why It Works:** Code documentation is largely formulaic—describe what the function does, what inputs it expects, what it returns, what errors it can raise. LLMs can generate this quickly and consistently.

**Pitfalls to Avoid:** Generated documentation can be vague, incorrect, or miss important details. Mitigate by having humans review and edit generated documentation, especially for complex or critical code.

### 4.3.8 Brainstorming and Planning

**The Use Case:** An engineer is designing a feature, considering architectural changes, planning a refactor, or working through a complex problem and wants to explore options, implications, and tradeoffs systematically.

**Where It Fits in D³:** DERIVE phase (exploring design options and planning), DISCOVER phase (understanding existing constraints and options).

**AI Roles:**

**Builder LLM:** Generates design proposals, explores tradeoffs, and articulates implications. "Here are three ways to implement this caching layer, with tradeoffs in terms of complexity, consistency guarantees, latency impact, and operational overhead."

**Reviewer LLM:** Critiques proposals, identifies risks, and raises alternative considerations. "You propose a write-through cache for consistency, but this increases write latency. Have you considered the impact on user-facing requests? What about using a write-behind cache with eventual consistency?"

**Why It Works:** Brainstorming benefits from exploring multiple perspectives and having a partner to argue against. LLMs can play this role effectively, proposing options systematically and helping engineers think through implications they might miss.

**Pitfalls to Avoid:** LLMs can propose plausible-sounding designs that have subtle flaws or that don't align with the organization's constraints or technical landscape. Mitigate by grounding design discussions in organizational context, by having domain experts review proposals, and by being explicit about constraints before brainstorming.

### 4.3.9 Initial Feature Scaffolding

**The Use Case:** An engineer is starting a new feature or module and needs help with the initial structure—class hierarchy, function signatures, module organization, overall architecture outline.

**Where It Fits in D³:** DERIVE phase (planning the work) and early DELIVER phase (bootstrapping implementation).

**AI Roles:**

**Builder LLM:** Creates code outlines, project structures, and skeleton implementations. Given a feature description, generates the initial class structure, file organization, and function signatures that the engineer can then flesh out.

**Reviewer LLM:** Examines the scaffolding for completeness, consistency, and alignment with architectural patterns used elsewhere in the codebase. "You created a service class for user management, but I notice the codebase uses a repository pattern elsewhere. Should this follow the same pattern for consistency?"

**Why It Works:** Starting a new feature is often the hardest part—deciding how to organize the code, what classes and interfaces to create, what the overall shape will be. LLMs can rapidly generate a reasonable scaffold that the engineer can refine, shortening the setup phase.

**Pitfalls to Avoid:** Generated scaffolding might not align with the codebase's architecture or conventions. Mitigate by providing the LLM with examples of how similar features are organized elsewhere in the codebase, and by having the engineer review and adjust the scaffold before fleshing it out.

### 4.3.10 Code Explanation

**The Use Case:** An engineer encounters unfamiliar code and needs to understand what it does, how it works, and why it was written that way—without needing to hunt down the original author or spend hours reading code and documentation.

**Where It Fits in D³:** DISCOVER phase (understanding existing systems), DELIVER phase (learning unfamiliar frameworks or patterns).

**AI Roles:**

**Builder LLM:** Reads code and provides clear, natural-language explanations of functionality. Can explain what each section does, what the overall flow is, and what the code is trying to achieve.

**Reviewer LLM:** Verifies that explanations are accurate and complete. "Your explanation of what this function does is clear, but you didn't mention that it uses reflection to dynamically load classes. That's an important detail for understanding performance characteristics."

**Why It Works:** Explaining code is a natural task for LLMs trained on public code repositories and documentation. Having an interactive partner to ask "what does this do?" is far faster than manually reading code or searching for documentation.

**Pitfalls to Avoid:** Explanations can be incorrect or misleading, especially for obscure patterns or non-standard code. Mitigate by asking the LLM to provide concrete examples and by verifying explanations against actual behavior.

---

# 5. Measuring AI Adoption and Impact

## 5.1 Surveys and Self-Reported Data

Organizations adopting AI need mechanisms to understand how well adoption is proceeding and what impact engineers perceive. Surveys are a direct, low-cost method to gather this signal. However, surveys must be carefully designed to capture meaningful information rather than just binary adoption metrics. While I acknowledge that self-reported metrics can diverge from objective velocity (Becker et al.), in brownfield environments, cognitive load and developer confidence are leading indicators of long-term maintainability, often more predictive than short-term speed. Appendix B provides a survey example.

**Survey Administration and Analysis**: Conduct surveys quarterly or semi-annually to track trends. Segment results by role (junior engineer, senior engineer, tech lead), team, and product area to identify pockets of high adoption and resistance. Use open-ended questions to capture nuanced feedback: "What is one thing that improved in your workflow with AI assistance? What is one challenge?" This qualitative data is often as valuable as quantitative scores.

**Limitations and Calibration**: Self-reported productivity and quality are subject to bias. An engineer who uses AI frequently might overestimate productivity gains (halo effect) or underestimate them (skepticism). To calibrate, combine survey data with objective metrics (described below). Also, be cautious of response bias—engineers who find AI tools valuable are more likely to respond to surveys about AI adoption, potentially inflating reported adoption rates.

## 5.2 Capacity Planning and Productivity Modeling

Beyond surveys, organizations can estimate productivity impact by modeling how AI affects team capacity. The idea is straightforward: if AI reduces the time spent on certain tasks, that time is freed for other work, increasing overall team capacity or enabling the same output with fewer people.

**Capacity Modeling Approach**:

1. **Identify Major Task Categories**: For a given team or organization, what does an engineer's time go to? Typical categories include:

    - Writing new features and business logic
    - Maintaining and fixing bugs
    - Debugging and incident response
    - Code review and mentorship
    - Documentation and knowledge work
    - Meetings and coordination
    - Learning and professional development
    - Administrative tasks

2. **Estimate Time Allocation (Baseline)**: Using surveys, time-tracking data, or estimation from managers and leads, establish a baseline: on average, engineers spend X% of their time on feature development, Y% on bug fixes, Z% on code review, etc. A common pattern is approximately 40–50% on new features, 20–30% on maintenance and bugs, 15–20% on code review and coordination, 10–15% on documentation, and 5–10% on other tasks.

3. **Identify High-Impact AI Opportunities**: Based on the D³ framework and evidence from surveys and research, which tasks benefit most from AI? Common high-impact areas:

    - Test generation: reduces time spent writing tests by 30–50%.

- Documentation: reduces time spent writing and updating documentation by 40–60%.
- Debugging: reduces time spent on initial diagnosis by 30–50%.
- Code review support: reduces senior engineers' review time by 20–30%.
- Onboarding and code understanding: reduces time spent learning unfamiliar systems by 20–40%.

4. **Estimate Productivity Gains**: For each high-impact area, estimate the percentage time savings and the percentage of a typical engineer's time devoted to that task. For instance:

   - If test generation saves 40% of the time spent on writing tests, and engineers spend 8% of their time writing tests, the net time savings is approximately 3.2% of total time.
   - If documentation assistance saves 50% of the time spent on documentation, and engineers spend 10% of their time on documentation, the net time savings is approximately 5% of total time.

5. **Aggregate Capacity Impact**: Sum the time savings across all high-impact areas to estimate total capacity impact. If the scenarios above hold, the team has gained approximately 8.2% additional capacity. For a 10-person team, this is equivalent to nearly one full-time person's worth of additional capacity.

6. **Map Capacity to Business Outcomes**: With additional capacity, what can the team accomplish? Options include:

   - Maintain current output with fewer people (1 FTE reduction).
   - Increase output by ~8% with the same team size.
   - Reduce on-call burden by allowing more time for systemic improvements.
   - Increase time for professional development and learning.

**Limitations and Caveats**: This model is necessarily approximate and should be calibrated with actual data. Not all engineers benefit equally from AI (experience level, task type, and tool familiarity all matter). Productivity gains may be front-loaded (initial enthusiasm) or subject to diminishing returns over time (as engineers move through low-hanging fruit tasks). Additionally, realizing capacity gains requires deliberate choice—if an organization doesn't redirect freed time to high-value work, the benefits are lost.

**Integration with D³**: The D³ framework makes capacity planning more predictable because work is decomposed into clear tasks with estimated effort. By tracking actual effort vs. estimated effort for AI-assisted vs. human-only work, an organization can empirically calibrate its capacity model over time, making future planning more accurate.

## 5.3 Telemetry and Workflow Analytics

Objective signals from tooling provide the most reliable data on AI adoption and impact. Organizations should instrument their development environment to collect telemetry on AI tool usage and outcome metrics.

**Usage Telemetry**:

- **Adoption Metrics**: How many developers are using AI tools? How frequently? (Daily, weekly, monthly active users.)
- **Task-Specific Usage**: What fraction of code review happens with AI assistance? What fraction of pull requests include AI-generated code? What fraction of documentation is generated or refined with AI?
- **Acceptance Rates**: When AI tools suggest code or documentation changes, what percentage does the engineer accept as-is, edit, or reject? High acceptance rates suggest the tool is providing high-quality suggestions; low acceptance rates suggest the opposite.
- **Feature Usage**: Which AI features are used most? (Code generation, documentation, explanation, refactoring suggestions, test generation, etc.) This indicates where the tool provides value.

**Outcome Metrics**:

- **Code Quality**: Do AI-assisted pull requests have comparable defect rates to human-written code? This can be measured through post-deployment defect rates, code review comment counts, or automated static analysis tools.
- **Review Metrics**: How does time-to-code-review change with AI assistance? Are pull requests with AI-generated code reviewed faster or slower? Do they receive more comments?
- **Pull Request Patterns**: Distribution of PR size, time-to-merge, and number of review rounds. Are PRs becoming smaller and faster to merge (positive signal) or larger and slower (potentially negative signal if quality is declining)?
- **Incident and Bug Rates**: Do teams using AI tools more extensively have higher or lower bug escape rates? Correlation (not causation) can inform whether AI is helping or hurting quality.
- **Onboarding and Ramp-Up**: Do new engineers on teams using AI tools reach productivity faster? Measure by tracking task velocity and quality for new hires over time.
- **Documentation Completeness**: Percentage of modules, services, or APIs with documentation. Has AI assistance led to higher documentation coverage?

**Combining Signals**:

The most powerful insights come from combining multiple signals. For instance:

- High adoption + high acceptance rates + stable/improving code quality → AI tools are genuinely adding value.
- High adoption + low acceptance rates + declining code quality → something is wrong (poor integration, tool not suited to codebase, engineers not trained).
- High adoption + stable quality but much faster PR merges → AI is helping with velocity without compromising quality.
- High adoption + slightly declining quality but significantly improved onboarding → tradeoff may be worth it (invest in code review practices to maintain quality).

**Avoiding Misinterpretation**:

Be careful not to draw premature conclusions from raw metrics. For instance:

- Faster PR merges could indicate improved quality or reduced review rigor. Combine with defect rate data to interpret.
- Lower documentation coverage after AI adoption could mean AI-generated docs are not reaching the knowledge base, or it could mean the AI is concentrating documentation in new systems while legacy systems are forgotten.
- Higher review comment counts could indicate reviewers are being more thorough (good) or that code quality is declining (bad). Context matters.

**Longitudinal Tracking and Feedback Loops**:

Establish baseline metrics before AI adoption, then track them quarterly or semi-annually. Look for trends over time rather than snapshot measurements. Use this data to inform decisions about where to invest more in AI support (high-value areas with room for improvement), where to reduce reliance (low-value areas), and what training or process changes might help.

---

# 6. Risks, Ethics, and Guardrails

While AI offers significant productivity potential, responsible adoption requires acknowledging and mitigating risks. Large organizations must establish clear governance, policies, and practices to ensure AI tools enhance rather than undermine engineering practices.

## 6.1 Code Quality and Technical Debt

AI-generated code can introduce new forms of technical debt if not carefully managed. LLMs may generate syntactically correct code that violates architectural principles, introduces unnecessary complexity, or creates subtle bugs that are not caught by automated tests. Mitigation strategies include:

- **Mandatory Human Code Review**: AI-generated code must be reviewed by humans before merging. This is non-negotiable for production systems. Code review processes should be explicit about scrutinizing AI-generated code, with reviewers alert to common AI pitfalls (e.g., overly complex solutions, missing edge case handling).
- **Coding Standards and Linting**: Establish clear coding standards and use automated tools (linters, static analysis) to catch common issues. AI tools should be configured to respect these standards, and any AI-generated code that violates them should be flagged for manual review.
- **Testing Requirements**: Comprehensive testing is critical. Organizations should require high test coverage for AI-assisted code and use mutation testing to verify that tests are meaningful. Property-based testing can catch subtle bugs in AI-generated code.
- **Review Gatekeeping**: For critical systems (payment processing, authentication, security-sensitive code), add an additional review gate—for instance, architecture review or security review—to catch higher-level issues that unit testing might miss.

## 6.2 Over-Reliance and Skill Atrophy

There is a risk that developers, particularly junior engineers, become over-reliant on AI tools and lose the ability to write code or diagnose problems without assistance. This is a legitimate concern, but it can be managed:

- **Deliberate Practice Without AI**: Encourage engineers to periodically work without AI assistance, particularly on learning tasks. Junior engineers should spend some portion of their time (e.g., 20–30%) solving problems independently to build foundational skills.
- **Teaching and Mentorship**: Use AI tools to accelerate task completion, but not to eliminate the need for senior engineers to mentor and teach. Pair junior and senior engineers on tasks, with AI assistance available to both, creating opportunities for learning.
- **Measurement and Feedback**: Track whether junior engineers are progressing in skill as expected. If AI tools are preventing skill development, adjust practices—for instance, by increasing independent work or pairing opportunities.
- **Conscious Tool Limitations**: Make it clear to teams that AI tools have limitations and are not always correct. Encourage skepticism and verification, particularly for novel or complex work.

## 6.3 Security and Privacy of Prompts and Code

When engineers use cloud-based AI tools, they send code and prompts to external services. This raises privacy and security concerns:

- **Data Classification and Policies**: Establish clear policies on what data can be sent to external AI services. Highly sensitive data (security vulnerabilities, proprietary algorithms, customer data) should not be shared with external tools. For sensitive work, use self-hosted or private LLM instances, or use vendor services with privacy guarantees (e.g., enterprise agreements with data retention policies).

- **Redaction and Anonymization**: When using external AI tools, have processes to redact or anonymize sensitive information in prompts. For instance, replace specific company names with placeholders, remove API keys or credentials, and avoid including customer data.
- **Audit and Compliance**: If the organization is subject to compliance requirements (HIPAA, PCI-DSS, GDPR, etc.), ensure that AI tool usage aligns with compliance obligations. This may require vendor assessments, data processing agreements, or restrictions on what work can use AI.
- **Vendor Management**: If using third-party AI services, establish vendor governance. Know where data is stored, how long it is retained, what the vendor's data usage policies are, and what happens if the vendor is breached or shuts down.

## 6.4 Licensing and Attribution

There is ongoing debate about whether code generated by LLMs trained on public code repositories (many of which are under open-source licenses) creates licensing obligations for the generated code. This is legally uncertain and evolving:

- **Organizational Policy**: Establish a clear policy on whether AI-generated code is treated the same as human-written code or whether it requires special handling. Some organizations accept generated code like any other code; others have restrictions (e.g., "use AI to generate code, but attribute it or restrict to compatible licenses").
- **Code Review Considerations**: During code review, reviewers should be alert to AI-generated code that closely resembles specific open-source projects, which could indicate licensing issues.
- **Vendor Responsibility**: Some AI vendors (e.g., GitHub with Copilot Business) offer licensing assurance or indemnification. Evaluate vendor terms and consider whether this risk-sharing is valuable for your organization.

## 6.5 Organizational Risk and Unequal Access

If AI tools are not evenly distributed across teams or organizations, they can exacerbate existing inequalities:

- **Access and Equity**: Ensure that all engineers have access to AI tools, not just a privileged subset. Unequal access can create two-tier teams where some engineers have productivity advantages and others do not.
- **Training and Enablement**: Provide training on effective use of AI tools. Without good training, some engineers will benefit more than others, again creating inequality.
- **Incentive Alignment**: Be cautious about performance metrics that might incentivize inappropriate AI use. For instance, if engineers are evaluated on code velocity, some might overuse AI in ways that reduce code quality. Use balanced metrics (velocity and quality, for instance) to avoid perverse incentives.

## 6.6 When NOT to Use AI

Despite the benefits, there are contexts where AI should not be used:

- **Safety-Critical Code**: Code where bugs could cause severe harm (medical devices, avionics, autonomous systems, power plant controls) should involve more conservative practices. AI-assisted development is acceptable if it includes more rigorous review and testing, but full automation is not appropriate.
- **Security-Sensitive Protocols**: Cryptographic code, authentication mechanisms, and security-critical algorithms should be reviewed by security experts, not delegated primarily to AI tools.
- **Novel Research**: If an engineer is exploring new research or unusual architectural approaches, AI tools trained on existing practices may not be helpful and could mislead.
- **Compliance and Legal Work**: Work subject to strict compliance requirements (regulatory filings, legal documentation, policy interpretation) should involve human domain experts, not AI generation.

## 6.7 Anecdotal Experience: The Experience Paradox in AI-Assisted Coding

Based on observations from leading AI adoption initiatives across enterprise and startup environments, a counterintuitive pattern emerged: junior engineers often derive less productivity benefit from LLM-generated code than their senior counterparts, despite widespread assumptions that AI tools would serve as "force multipliers" for less experienced developers.

This experience paradox stems from several interrelated factors. Senior engineers possess the pattern recognition and architectural intuition necessary to quickly evaluate whether LLM-generated code aligns with system constraints, coding standards, and performance requirements. They can distinguish superficially correct code from genuinely appropriate solutions—recognizing when an LLM has produced syntactically valid but architecturally problematic implementations. Their accumulated experience enables them to ask better questions, provide more precise context in prompts, and rapidly identify edge cases or failure modes that LLM outputs might miss.

Conversely, junior engineers face a validation problem: without deep experience, they lack the expertise to confidently assess code quality, appropriateness, or correctness. An LLM might generate code that compiles and passes basic tests but introduces subtle bugs, violates team conventions, creates technical debt, or misses critical security considerations. Junior engineers are more likely to accept such code uncritically or spend disproportionate time trying to debug AI-generated implementations they do not fully understand. Rather than accelerating learning, premature reliance on AI-generated code can actually slow skill development by obscuring fundamental patterns and problem-solving approaches that junior engineers need to internalize.

# 7. Conclusion

The software engineering industry stands at an inflection point. Large language models have demonstrated remarkable capabilities in coding and documentation, and adoption is accelerating across organizations of all sizes. However, the path from capability to sustained productivity gain is not automatic. Our findings suggest perceived benefits in productivity, cognitive load, and code quality, but these results should be interpreted as initial evidence of practitioner sentiment rather than validated outcomes. This paper has argued that organizations, particularly large enterprises managing complex brownfield systems can unlock material productivity benefits by adopting a disciplined approach anchored in the **Discover–Define–Deliver (D³) AI Productivity Framework**.

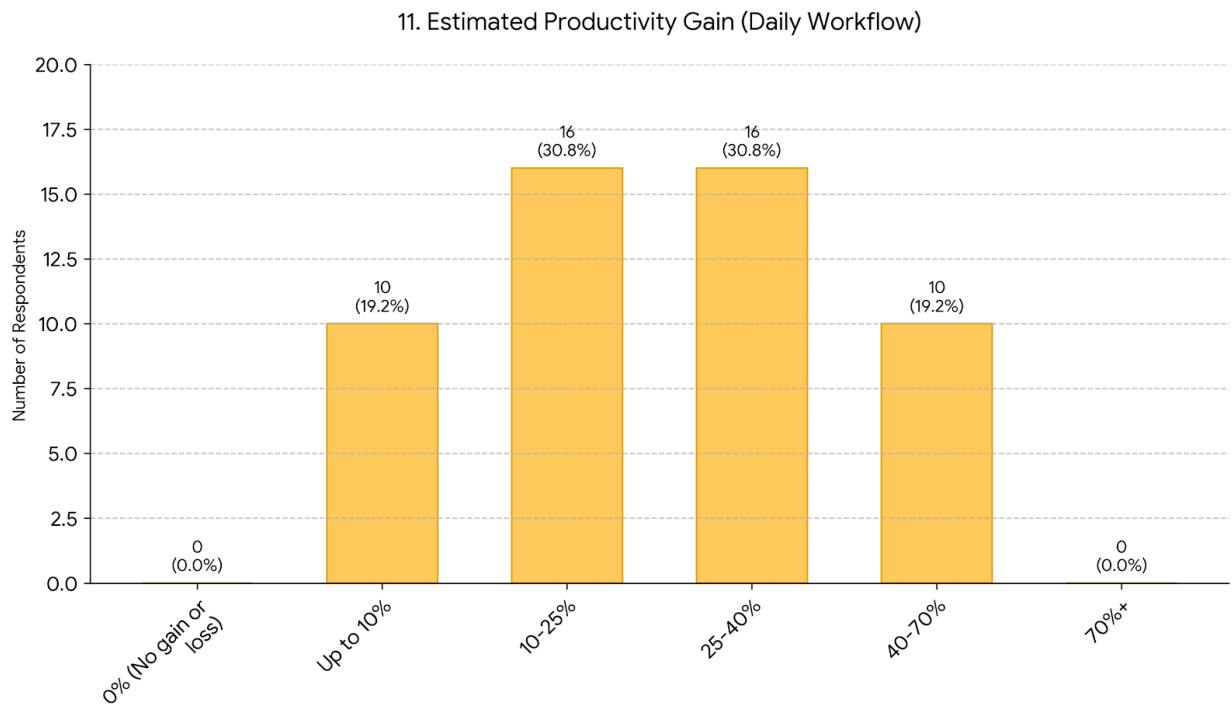

Fig. 6: As per the survey results, the **weighted average productivity** gain is approximately **26.9%** and **average productivit**y gain is between **25% to 40%**.

The core thesis is that **AI is most powerful when it reduces cognitive load while preserving human control**. Rather than attempting to automate away the need for human engineering judgment, the D³ framework uses AI to augment human decision-making across three phases:

- **DISCOVER**: AI helps engineers rapidly build comprehensive understanding of existing systems, reducing the time spent reading code and reconstructing institutional knowledge.

- **DEFINE**: AI helps engineers design complex changes, identify risks, and decompose work into parallelizable tasks, enabling teams to move with confidence even in the face of architectural complexity.
- **DELIVER**: AI assists in code generation and pre-review, providing fast feedback that accelerates the development cycle while preserving human review and judgment as mandatory gates.

The dual-agent architecture pairing a **Builder LLM** for generation with a **Reviewer LLM** for critique, creates a form of discipline that improves outcomes. The Builder LLM is optimized for coverage and rapid hypothesis generation; the Reviewer LLM is optimized for rigor and critical thinking. Together, they create a workflow that is both productive and thoughtful.

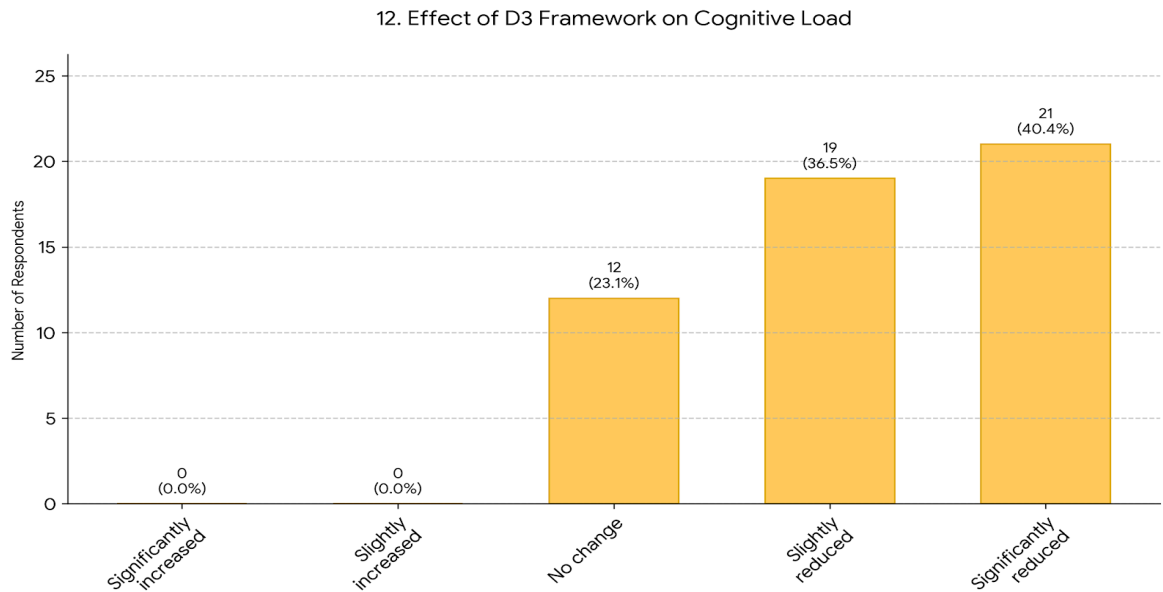

Fig. 7: Survey results highlight that the D3 framework reduces cognitive load by 77%.

However, realizing these gains requires deliberate effort. Organizations must:

1. **Invest in Methodology and Training**: The D³ framework is not intuitive; teams need training and mentoring to apply it effectively. This includes both technical training (how to use specific AI tools effectively) and process training (how to integrate AI into team workflows).
2. **Establish Clear Governance and Policies**: Policies around data security, code quality standards, review requirements, and appropriate use cases provide guardrails that prevent problems and build confidence.
3. **Measure and Iterate**: Implement the measurement approaches outlined in this paper surveys, capacity modeling, and telemetry to understand adoption and impact. Use this data to refine practices over time.
4. **Balance Productivity and Sustainability**: While the goal is to increase productivity, organizations must be cautious not to sacrifice long-term sustainability and developer wellbeing. AI tools should free engineers to focus on meaningful work, not increase workload through constant pressure to do more.

Looking forward, several promising directions merit attention:

- **Deeper Multimodal Integration**: As LLMs become more sophisticated in reasoning about images, diagrams, and other modalities, opportunities to assist with architecture understanding, debugging with visual log analysis, and UI/backend integration will expand.
- **Closer Integration with Observability**: Tightly coupling AI tools with observability systems (logs, metrics, traces) enables more powerful debugging and diagnosis. An LLM that can reason about production behavior, correlate traces across services, and propose root causes will be far more valuable than one working only from code.
- **Rigorous Empirical Evaluation**: Most current research on AI productivity is limited in scope. We need larger-scale, longitudinal studies examining how sustained use of AI affects productivity, code quality, and developer satisfaction in diverse organizational contexts and brownfield environments.
- **Human-AI Collaboration Interfaces**: As understanding of effective AI-human workflows improves, better interfaces and tools will emerge that reduce friction and amplify collaboration.

For senior engineers, tech leads, and engineering executives at large organizations, the message is clear: AI-driven productivity is achievable, but it requires thoughtful integration into discipline processes. Those that treat AI as a magical tool requiring no process or discipline will face disappointment and risk.

## Acknowledgements

This work synthesizes insights from recent academic research, industry surveys, and practical experience with generative AI tools in large software engineering organizations. I am grateful to the 52 software engineering professionals who participated in the survey study and generously shared their experiences applying the D³ framework to real-world brownfield systems. Their candid feedback made this research possible. I also acknowledge the engineering teams across multiple companies who shared their experiences with AI tools in legacy systems, allowing this framework to be grounded in practical experience rather than theory alone. I acknowledge the growing community of practitioners and researchers focused on responsible AI adoption in software engineering. Their work on governance, measurement, and risk mitigation has informed the ethical and practical guidelines outlined in this paper. I extend my thanks to Amit Mishra (Amazon), Praveen Kumar (Amazon) and other engineering leaders from various organizations for their insights. I also want to thank the startups (NuoFintech, Kirtanmantra) who were open to adopting the D³ framework. Finally, I recognize that this paper represents a snapshot of a rapidly evolving field. As AI capabilities mature and organizational practices evolve, the frameworks and guidance presented here will require refinement. I welcome feedback and case studies from practitioners implementing these ideas in their own organizations.

**Declaration of Generative AI and AI-Assisted Technologies in the Writing Process**

During the preparation of this work, I used **Gemini 3 Pro, ChatGPT-5.1, and Claude Sonnet 4.5** acting as both content creators and reviewers. I utilized these tools to assist with drafting initial outlines, brainstorming structure, fixing grammar, summarizing data, refining prose for clarity, and formatting citations. In doing so, I applied the principles of the D³ framework introduced in this paper to the writing process itself. Additionally, the specific coding prompts and example outputs presented in Section 4 were generated with the assistance of these tools to illustrate the framework. I reviewed and edited all AI-generated content for accuracy and readability, and I take full responsibility for the final content of this publication.

"AI Developer Survey." *2025 Stack Overflow Developer Survey*, Stack Overflow, 2025, [survey.stackoverflow.co/2025/ai](survey.stackoverflow.co/2025/ai).

Sun, Tao, et al. "BitsAI-CR: Automated Code Review via LLM in Practice." *Proceedings of the 33rd ACM International Conference on the Foundations of Software Engineering*, 2025. [arxiv.org/pdf/2501.15134](arxiv.org/pdf/2501.15134).

"The 2024 State of Developer Productivity." *Cortex*, 2024, [cortex.io/report/the-2024-state-of-developer-productivity](cortex.io/report/the-2024-state-of-developer-productivity).

"The 90-Day Developer Onboarding Best Practices That Work." *FullScale*, 2025, [fullscale.io/blog/developer-onboarding-best-practices](fullscale.io/blog/developer-onboarding-best-practices).

Ziegler, Albert, et al. "Measuring GitHub Copilot's Impact on Productivity." *Communications of the ACM*, vol. 67, no. 3, Mar. 2024, pp. 54-63. *ACM Digital Library*, [doi.org/10.1145/3633453](doi.org/10.1145/3633453)

---

# Appendix:

## Appendix A: Survey results on AI Productivity and D3 framework

**Total Survey Respondents:** 52

### Section 1: Demographics & Context

**1. How many years of professional software engineering experience do you have?**

- 1-3 years: **6**
- 3-6 years: **15**
- 6-10 years: **21**
- 10+ years: **10**

**2. What is the approximate size of your engineering organization?**

- 1-500 engineers (Startup/Small): **8**
- 500-3,000 engineers (Mid-Market/Growth): **11**
- 3,000+ engineers (Enterprise/Big Tech): **33**

**3. Which best describes the codebase you primarily work on?**

- Greenfield (New projects, modern stack): **8**
- Brownfield (Legacy systems, technical debt): **27**
- Mixed (Even split): **17**

## Section 2: AI Usage & Baseline

**4. Which AI models or tools do you use for your daily work?**

- Anthropic Claude: **32**
- ChatGPT: **10**
- Google Gemini Pro: **10**

**5. How would you rate the usefulness of AI when used "Ad-Hoc" (e.g., simple chat or autocomplete without a structured workflow)?**

- 1 - Not useful: **13**
- 2 - Slightly useful: **22**
- 3 - Moderately useful: **15**
- 4 - Very useful: **2**
- 5 - Transformative: **0**

## Section 3: Evaluating the D3 Methodology

**6. When using a structured framework where AI is used to "Explore & Explain" legacy code (The "Discover" Phase), how does it impact your understanding?**

- It adds unnecessary overhead/time: **3**
- No significant difference: **5**
- It helps slightly with context: **14**
- It significantly reduces time spent reading code/docs: **24**
- It provides "Senior Engineer" level context instantly: **6**

**7. When using AI to "Plan & Architect" a solution before writing code (The "Define" Phase), how does it affect your rework rate?**

- I end up rewriting more code: **1**
- No change in rework: **7**
- Slightly less rework: **13**
- Significantly less rework (catches design flaws early): **25**
- Drastic reduction (rarely hit architectural dead-ends): **6**

**8. Have you utilized a "Reviewer Agent" (a separate AI prompt/agent tasked specifically with critiquing your code/plan)?**

- No, I only use AI to generate/build: **20**

- Yes, but rarely: **18**
- Yes, frequently: **14**

**9. If "Yes" to Q8: How effective is the "Reviewer Agent" compared to human code review for catching basic defects?** *(Answered by the 44 respondents who used the agent)*

- Worse than a human: **0**
- Comparable to a junior peer: **14**
- Comparable to a senior peer: **10**
- Better than a human at catching subtle/logical bugs: **8**

**10. How would you rate the usefulness of AI when applied via the D3 Framework (or similar structured workflows) compared to Ad-Hoc usage?**

- 1 - Much less useful: **0**
- 2 - Slightly less useful: **3**
- 3 - About the same: **10**
- 4 - More useful (better quality/reliability): **26**
- 5 - Significantly more useful: **13**

## Section 4: Impact & Outcomes

**11. Estimated overall productivity gain in your daily workflows using AI for documentation and Coding:**

- 0% (No gain or loss): **0**
- Up to 10%: **10**
- 10-25%: **16**
- 25-40%: **16**
- 40-70%: **10**
- 70%+: **0**

**12. How has the use of the D3 framework affected your "Cognitive Load" (mental effort/exhaustion) when working on complex legacy code?**

- Significantly increased load: **0**
- Slightly increased load: **0**
- No change: **12**
- Slightly reduced load: **19**
- Significantly reduced load (I feel less drained): **21**

# Appendix B: Example Survey Instrument for AI Adoption and Impact

**Quarterly Developer AI Adoption Survey**

**Introduction**: We are measuring the adoption and impact of AI tools in our engineering organization. Your feedback is valuable and confidential. Survey takes ~5 minutes.

**Section 1: Usage**

1. In the past month, how often have you used AI coding or documentation tools (Copilot, Claude, ChatGPT, etc.)?

    - Never
    - Rarely (< 1 hour per week)
    - Sometimes (1–5 hours per week)
    - Frequently (5–20 hours per week)
    - Very frequently (> 20 hours per week)

2. For the following tasks, how much do you use AI assistance? (Scale: Never, Rarely, Sometimes, Often, Always)

    - Writing unit tests: ___
    - Generating code: ___
    - Generating documentation/comments: ___
    - Debugging errors: ___
    - Code review and quality checks: ___
    - Refactoring legacy code: ___
    - Learning unfamiliar frameworks/code: ___
    - Designing architecture: ___

3. Which AI tools do you use most frequently? (Select all that apply)

- GitHub Copilot
- ChatGPT
- Claude
- Gemini or other Google tools
- Our internal AI assistant
- Other: ___

**Section 2: Perceived Impact**

4. How has AI assistance affected your productivity?
   - Significantly decreased
   - Somewhat decreased
   - No effect
   - Somewhat increased
   - Significantly increased

5. How has AI assistance affected the quality of code you write?
   - Significantly worsened
   - Somewhat worsened
   - No effect
   - Somewhat improved
   - Significantly improved

6. How confident are you in code generated by AI before human review?
   - Very low (almost always needs significant revision)
   - Low (often needs revision)

- Neutral
- High (usually works with minor edits)
- Very high (usually works as-is)

7. Rate your agreement with the following statements. (Scale: Strongly Disagree, Disagree, Neutral, Agree, Strongly Agree)
    - AI tools have reduced the time I spend on routine tasks: ___
    - AI tools have helped me learn new technologies faster: ___
    - AI tools make documentation easier to write and maintain: ___
    - I trust human-written code more than AI-generated code: ___
    - AI tools have improved my job satisfaction: ___
    - I would use AI tools more with better training: ___

**Section 3: Barriers and Needs**

8. What is the primary barrier to using AI tools more?
    - Lack of training or familiarity
    - Concerns about code quality
    - Privacy or security concerns
    - Tool not integrated well into my workflow
    - Uncertain when AI adds value
    - Other: ___

9. What would help you use AI tools more effectively?
    - Better training or documentation
    - Better integration with my IDE

- Clearer guidelines on appropriate use cases
- Assurance about code quality and security
- Better tool performance or fewer false positives
- Other: ___

**Section 4: Open Feedback**

10. What is one way AI tools have improved your work?

11. What is one challenge you face with AI tools?

12. Any other feedback or suggestions?